\documentclass{aa}
\usepackage[varg]{txfonts}
\usepackage{natbib}
\bibpunct{(}{)}{;}{a}{}{,} 
\usepackage{subfig}
\usepackage{graphicx}

\newcommand{\um}{$\mu$m}
\newcommand{\kms}{\mbox{km\,s$^{-1}$}}
\newcommand{\Msun}{$M_{\odot}$}

\newcommand{\cms}{${\rm cm}^{-2}$}
\newcommand{\cmc}{${\rm cm}^{-3}$}
\newcommand{\hh}{H$_{2}$}

\newcommand{\ammonia}{NH$_{3}$}
\newcommand{\submm}{submm}

\newcommand{\ntwohp}{N$_2$H$^+$}

\newcommand{\hto}{H$_2$O}
\newcommand{\htoiso}{H$_2^{18}$O}
\newcommand{\csto}{C$^{17}$O}
\newcommand{\meth}{CH$_3$OH}
\newcommand{\irdcone}{G28.34+0.06}
\newcommand{\irdctwo}{G11.11-0.12}
\newcommand{\onea}{G28-NH$_3$}
\newcommand{\oneb}{G28-MM}
\newcommand{\twoa}{G11-NH$_3$}
\newcommand{\twob}{G11-MM}
\newcommand{\ctfs}{C$^{34}$S}
\newcommand{\cch}{CCH}
\newcommand{\scm}{cm$^{-2}$}
\newcommand{\ccm}{cm$^{-3}$}

\begin{document}

\title{\bf \rm  The \rm physical conditions in IRDC \bf \rm  clumps \rm from \bf \rm  Herschel \rm HIFI observations of H$_2$O \thanks{ {\it Herschel} is an ESA space observatory with science instruments provided by European-led Principal Investigator consortia with important participation of NASA.}}

\titlerunning{The physical conditions in IRDC clumps}

\author{R.F. Shipman \inst{1,2} 
          \and F.F.S van der Tak\inst{1,2}
          \and F.  Wyrowski \inst{3}  
           \and F. Herpin \inst{4,5}   
           \and W. Frieswijk \inst{6}
	  }

\institute{SRON Netherlands Institute for Space Research
   \and Kapteyn Astronomical Institute, University of Groningen
   \and Max-Planck-Institut f\"ur Radioastronomie 
   \and Univ. Bordeaux, LAB, UMR 5804, F-33270, Floirac, France
   \and CNRS, LAB, UMR 5804, F-33270, Floirac, France
      \and ASTRON Netherlands Institute for Radio Astronomy}

\date{}

\abstract {The earliest phases of high-mass star formation are poorly understood.}
{Our goal is to determine the physical conditions and kinematic structure of massive  \bf \rm  starforming \rm cloud \bf \rm  clumps \rm.}
{We analyse \hto\ 557 GHz line profiles observed with HIFI toward four positions in two infrared-dark cloud \bf \rm  clumps \rm.  By comparison with ground-based \csto, \ntwohp, \meth, and \ammonia\ line observations, we constrain the volume density and kinetic temperature of the gas and  estimate the column density and abundance of \hto\ and \ntwohp.}
{The observed water lines are complex with emission and absorption components.  The absorption is redshifted and consistent with a cold envelope, while the emission is interpreted as resulting from protostellar outflows.  \bf \rm  The gas density in the clumps is $\sim10^7$ \ccm.  The o-\hto\ outflow column density is $0.3-3.0 \times 10^{14}$ \scm.  The o-\hto\ absorption column density is between $1.5\times 10^{14}$ and  $2.6 \times 10^{15}$ \scm\ with cold o-\hto\ abundances between $1.5 \times 10^{-9}$ and  $3.1 \times 10^{-8}$.}
{All \bf \rm  clumps \rm have high gas densities ($\sim10^7$ \ccm) and display infalling gas. Three of the four clumps have outflows.  The \bf \rm  clumps \rm form an evolutionary sequence as probed by \hto\,  \ntwohp, \ammonia, and \meth.  \bf \rm  We find that \oneb\ is the most evolved, followed by \twob\ and then \onea.  The least evolved clump is \twoa\ which shows no signposts of starformation; \twoa\ is a high-mass pre-stellar core. }
\keywords{ISM, Star Formation}
\maketitle

\section{Introduction \label{introduction}}

The earliest stages of high-mass star formation are challenging to investigate observationally.  \bf \rm  Massive stars are rare and typically at large distances and they    evolve rapidly during their lifetimes  \citep[][]{mottram2011} .   It is expected that the very beginning of formation is also rapid. In addition, the high extinction and crowded nature of high-mass starforming regions complicate their study\rm.  Nevertheless, the formation of high-mass stars is of interest because of their large impact on the Galactic environment \citep[][]{tan2014}.

Since the late 1990s, the investigation into the very earliest stages has undergone significant growth, in a large part because of the discovery of the so-called infra-red dark clouds (IRDCs) \citep[][]{beuther2007}. 
These dark clouds were originally noted in \bf \rm \it{ Infrared Space Observatory} \rm \citep[][]{perault1996} and \bf \rm \it{ Midcourse Space eXperiment} \rm \citep[][]{egan1998} mid-infrared observations of the Galactic Plane.  They are easily seen as isolated dark filamentary patches or silhouettes  against the bright Galactic Plane.  To be optically thick in the mid IR implies high column densities of gas and dust \citep[$10^{22}$ cm$^{-2}$;][]{carey1998}.  These clouds were not seen in IRAS 100 \um\ data, and therefore are likely colder than 20 K \citep[][]{egan1998} and at distances of  a few to 10 kpc; IRDCs are therefore cold, dense, and massive clouds. 

Observations at \submm\ wavelengths (both molecular lines and continuum) not only confirmed the cold, dense, and massive nature of the clouds \citep[][]{carey2000,pillai2006a,pillai2006b},  but also identified isolated \bf \rm  clumps \rm within the IRDC filaments \bf \rm  \citep[][]{johnstone2003, rathborne2006}\rm.   Further studies \bf \rm  reveal that within these clumps  are cores in the process of starformation \citep[][and references therein]{ragan2012..herschel, henning2010, beuther2007}.  The cores often show molecular outflows and/or masers similar to other well-known massive starforming regions. \rm   The one main difference is that the forming sources have not yet had time to fully heat up and disrupt their environment \citep{beuther2007}. \footnote{ \bf \rm  This paper uses the term "clump"  to identify a part of the IRDC which has an enhancement of either \submm\ or \ammonia\ emission.  For approximate scales, IRDCs: 10 pc,  clumps: 0.5 pc, cores: 0.05 pc.  \rm} 

Recent investigations have focused on obtaining a chemical inventory within samples of IRDCs \citep[][]{miettinen2012,sanhueza2012} to try to identify trends and piece together evolution scenarios \citep[][]{chen2010}.  High resolution interferometric observations indicate that the starforming cores  are often made up of smaller more compact objects pointing to cluster formation as opposed to the formation of a single star \bf \rm  \citep[][]{rathborne2009, zhang2009, wang2011, wang2014}.  \rm

Observations of water transitions in \bf \rm  starforming \rm regions with the \it{Herschel Space Observatory} \rm have opened a new \bf \rm  means \rm of probing the processes occurring during star formation.   Water exposes many dynamic effects such as outflows, but water absorption also reveals many foreground clouds \citep{marseille2010}.   Water abundances are sensitive to temperature as the water sublimates from grain surfaces \citep{herpin2012} and show very complex structures within massive \bf \rm  starforming \rm \bf \rm  clumps \rm.


\bf \rm  The goal of this paper is to present the results of \it{Herschel}\rm \bf \rm-HIFI and \it{APEX} \rm \bf \rm observations  of a sample of IRDCs clumps at different evolutionary stages and to compare these results with other starforming regions observed by HIFI. \rm
This paper presents o-\hto\ 1$_{10} \to 1_{01}$  and \csto\ $ 3 \to 2 $  observations of four \bf \rm  starforming clumps \rm from two different clouds \irdcone\ and \irdctwo.  The active \bf \rm  clumps \rm are chosen based on their bright \submm\ continuum \citep{rathborne2006} while the quiescent clumps are chosen based on their prominent \ammonia\ line emission \citep{pillai2006b} The sources have been relatively well studied in the literature which allows our observations of the \bf \rm  clumps \rm to be placed into an existing  context.  

\begin{figure}
\includegraphics[width=9.0cm]{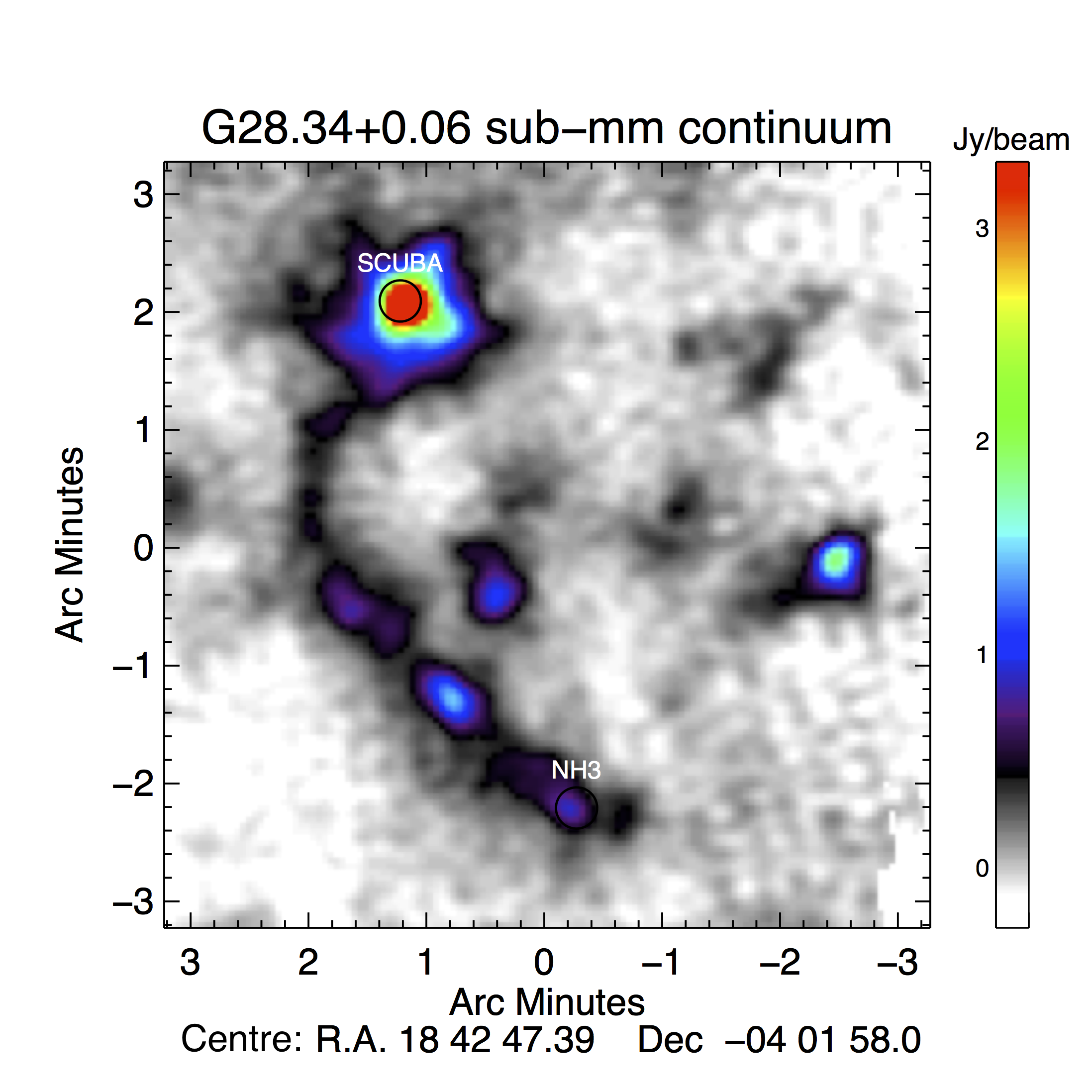}
\caption{SCUBA 850 $\mu$m image of  \irdcone.  The positions of the observed \bf \rm  clumps \rm are labelled as "SCUBA" for the \submm\ position and "NH3" for the \ammonia\ position.  The image is colour scaled from black (331 mJy/beam) to red (5951 mJy/beam).  Every change in colour is roughly an increase of 331 mJy/beam.   At a distance of 4.8 kpc, one arcminute is about 2 pc} \label{figirdcone}
\end{figure}

\begin{figure}
\includegraphics[width=9.0cm, angle=00]{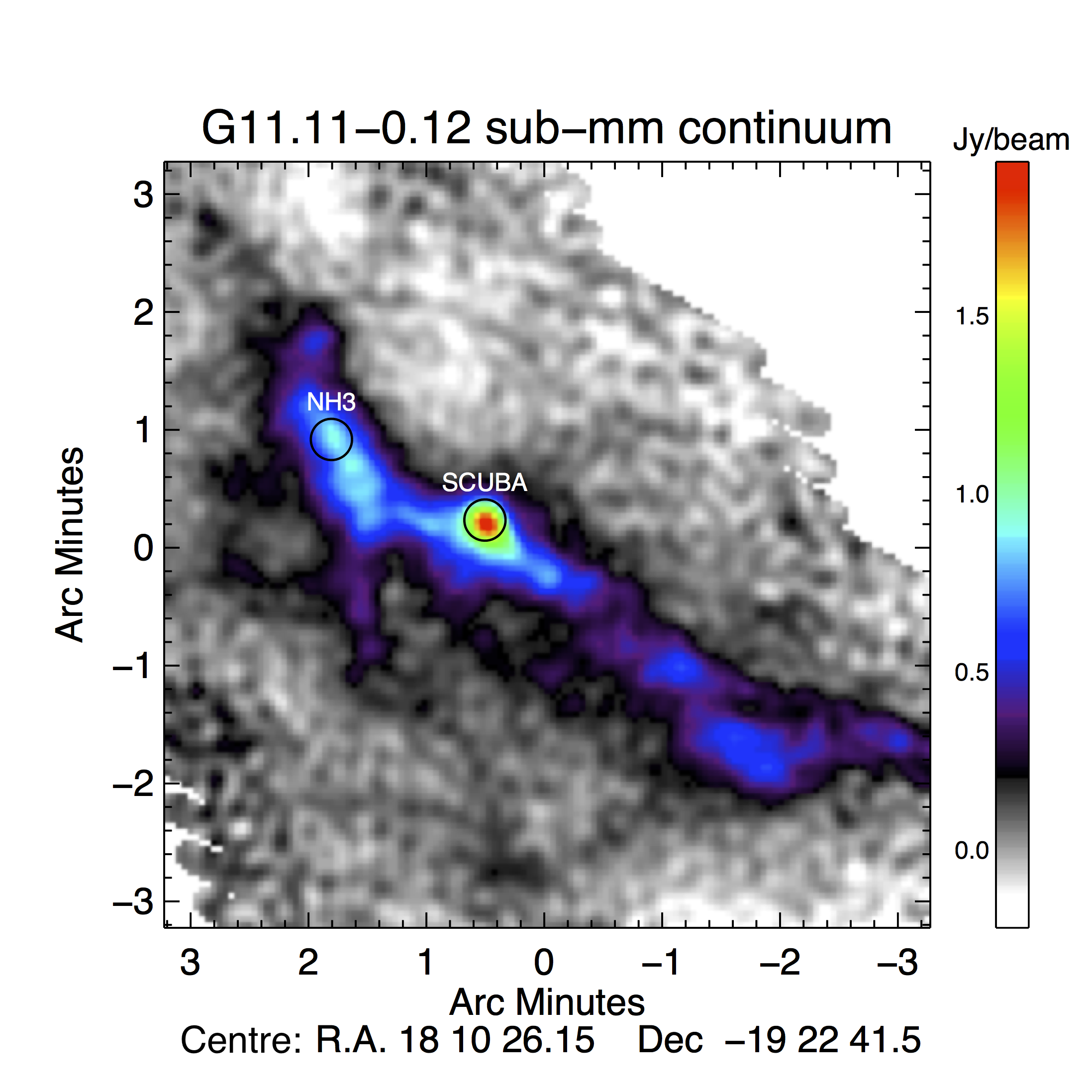}
\caption{SCUBA 850 $\mu$m image of \irdctwo. The positions of the observed \bf \rm  clumps \rm are labelled as "SCUBA" for the \submm\ position and "NH3" for the \ammonia\ position.   The image is colour scaled from black (290 mJy/beam) to red (1932 mJy/beam). Every change in colour is roughly an increase of 290 mJy/beam.    At the distance of 3.6 kpc, one arcminute is about 1.5 pc. } \label{figirdctwo}
\end{figure}

Figures ~\ref{figirdcone} and~\ref{figirdctwo} show 850 \um\ images of the two IRDCs \footnote{ Data from the  SCUBA  instrument on The James Clerk Maxwell Telescope are used for this investigation.  The JCMT is operated by the
Joint Astronomy Centre on behalf of
the Science and Technology Facilities Council of the United Kingdom,
the Netherlands Organisation for Scientific Research, and
the National Research Council of Canada.}.   The two clumps in \irdcone\ are labelled \onea\ for the position chosen based on \ammonia\ emission \citep{pillai2006b} and \oneb\ for the position chosen based on the strongest \submm\ emission \citep{rathborne2006}.   Likewise, two positions in  \irdctwo\ are labelled  \twoa\ \citep[\ammonia\ peak,][]{pillai2006b}   and \twob\   \citep[continuum peak,][]{rathborne2006}.  

Table \ref{objectsinfo} lists general clump properties.   Clump systemic velocity, gas temperature are based on \ammonia\ observations while clump mass and total column density are based on dust continuum emission \citep{carey1998,pillai2006a}.  \bf \rm  Kinematic distances to the clumps are taken from \citet[][]{carey1998} and are based the radial velocity of molecular line emissions and the galactic rotation curve. \rm The positions and Herschel observations are shown in Table \ref{objects}.
\bf \rm

The clumps span a range of evolutionary stages.   Both \oneb\ and \twob\ stand out as prominent point sources in PACS photometry \citep[][]{ragan2012..herschel}.  Both sources are also associated with \meth\ and/or \hto\ maser emission \citep[][]{pillai2006a, wang2008, wang2014} and both are seen to fragment into substructures at high spatial resolution \citep[][]{zhang2009, wang2014}.   A \meth\ outflow is observed for \twob\ \citep[][]{leurini2007, gomez2011, wang2014} and \oneb\ was found to harbour a "hot-core" \citet[][]{zhang2009}.  These clumps are active.

The clump \onea\ was studied together with \oneb\  and found to be of a younger evolutionary stage based on deuteration level \citep[][]{chen2010}.  Neither \ammonia\ clump is identified in \citet[][]{ragan2012..herschel}.  Based on their maps, \onea\ and \twoa\ are seen at SPIRE wavelengths but do not stand out as point sources at PACS wavelengths implying low temperatures ($<16$ K).  These clumps are not active.

\rm

\bf \rm   In Sect. \ref{allaboutdata} we give details of the observations and data reduction.  The line fitting procedure and results are presented in Sect. \ref{results}.   The column densities of \hto\ , \ntwohp\ , \csto\ , and \meth\ as well as gas densities based on \ntwohp\ and \meth\ transitions are discussed in  Sect. \ref{analysis}.  In Sect. \ref{discussion} the various dynamic structures in the clumps are discussed and an evolutionary sequence is proposed and the \hto\ results are compared with other starforming regions. \rm

\section{Observations \label{allaboutdata}}

\begin{table*}
\begin{center}
\caption{Gas properties of \bf \rm  clumps \rm \label{objectsinfo}}
\begin{tabular}{lcrccr}
\hline\hline
 ID&$v_{LSR}$\tablefootmark{1}& T$_{kin}$\tablefootmark{1}  &  $N$(H+H$_2$)\tablefootmark{1} &Distance\tablefootmark{2} &Mass\tablefootmark{1}  \\
     &  (\kms)   &     (K)          & ($10^{22}$ cm$^{-2}$)& (kpc)  & (\Msun )\\
\hline
\onea&80.2  & 13.2  &  5.9  & 4.8&  374 \\
\oneb&78.5  & 16.0  & 32.7 & 4.8 &2310\\
\twoa&30.4  & 12.7  &   8.4 & 3.6 &485 \\
\twob&29.2  &  13.8 & 10.2 & 3.6 & 172 \\

\hline
\end{tabular}
\tablefoot{References. (1) \citet[][]{pillai2006b}; (2) \citet[][]{carey1998}}

\end{center}
\end{table*}

\subsection{Herschel/HIFI observations}
The o-\hto\ ground state line observations were carried out as a part of the "Water In \bf \rm  starforming \rm regions with Herschel" key programme  \citep[WISH,][]{vandishoeck2011} with the HIFI instrument \citep{degraauw2010} on \it{Herschel} \rm  \citep{pilbratt2010}.
%

The observations were made with HIFI Band 1A using the double beam switch observing mode. 
The sideband separation of 8 GHz and IF bandwidth of 4 GHz allow a local oscillator (LO) setting where the o-\hto\ 1$_{10}$-1$_{01}$ transition at 556.9361 GHz, the \htoiso\ $1_{10}-1_{01}$ line at 547.676 GHz and the \ntwohp\ 6-5 transition at 558.96 GHz are observed simultaneously.   HIFI has two spectrometers, the acousto-optical Wide Band Spectrometer (WBS) and the correlator-based High Resolution Spectrometer (HRS).    The data from WBS  have a resolution of 1.1 MHz (0.5 \kms at 557 GHz) and the HRS was configured in two sub-bands centred on  the \hto\  and \htoiso\ lines with a resolution of 0.24 MHz (0.11 \kms).   The \ntwohp\ line was only observed with the WBS. 

The observations were started during the Science Demonstration phase of HIFI.   In these early observations, spurious signals due to an impure LO signal were still present at frequencies near the 557 GHz water line.   
The "water spur" affects mainly the very centre of the IF as the observations were set up.    In the earliest observations of \irdctwo\ the spurious responses hampered a proper observation of the \ntwohp\ and \htoiso\ lines.  These observations were repeated at a later date to improve the signal to noise and take advantage of the purified  LO chain.  Care was taken to make sure that data used in this study do not contain spurs.

The HIFI data were reduced and further analysis performed using the standard HIPE software\footnote{\bf \rm  Pipeline processing, data analysis and line fitting were carried out with the HIPE 10.0 software \citep{ott2010} \rm and the CASSIS plugin for HIPE and Cologne Database for Molecular Spectroscopy \citep[][ http://www.astro.uni-koeln/cdms]{cdms2005}.  CASSIS has been developed by IRAP-UPS/CNRS (http://cassis.irap.omp.eu). }.  The HIFI pipeline included updated beam efficiencies as listed in \citet{roelfsema2012}.     The vertical and horizontal polarization data have been inspected for standing waves and spurious signals separately.  Data from OBSID 134194462 have a low level standing wave which was removed using standard processing.   Finally, the double sideband data are corrected to single sideband by estimating and subtracting the emission from the image sideband.  This procedure is described in Appendix \ref{sidebandcorrection}.  The data shown in Fig. \ref{observedlineplots}  show single sideband data.

For the \ntwohp\ ($6-5$) line in G11.11, only OBSIDs 1342205490 and 1342206660  were used since spurious signals are present in the other observations on these sources.   Table \ref{objects} lists which line frequencies in which dataset are effected by spurs.
When  possible, both polarizations and multiple observations were co-added.   

The HRS data were used for the \htoiso\ lines and smoothed to 1.1 MHz resolution.  HRS data were also used for the \hto\ lines for \twoa\ and \twob.   The HRS band was not broad enough for the \oneb\ \hto\ emission and the \hto\ line was placed too close to the edge of the HRS band to use for \onea.   When smoothed, the HRS gives consistent results with the WBS.

\begin{table*}
\begin{center}
\caption{Overview of observations \label{objects}}

\begin{tabular}{llccccllcc}

\hline\hline

Short name & Long Name \tablefootmark{a} & OD\tablefootmark{b} &  OBSID \tablefootmark{c}      & Spur\tablefootmark{d}     &        & RA\tablefootmark{e} & DEC\tablefootmark{e}  & $\Delta$(RA)\tablefootmark{f} & $\Delta$(Dec)\tablefootmark{f}               \\
   &    &  &   &   &    & (J2000) & (J2000)  & (\arcsec)& (\arcsec)              \\
\hline
\onea&G28.34+0.06 P3 &332& 1342194461 & \ntwohp  &      & 18 42 46.40 &$-$4  04  12.0  &0.007&-0.742 \\  
      &&  707   &    1342219195    &    &              &                         &         & -0.051 & -0.75            \\
 \oneb& G28.34+0.06 P2 &332&1342194462 &  &       & 18 42 52.40 &$-$3 59 54.0  & -1.11  &-1.01\\
      &&   694  &  1342218945   &  &              &                         &                   & -1.098 &-1.00	   \\
 \twoa&  G11.11-0.12 P1 &291&  1342191497  &  \ntwohp &      & 18 10 33.90 &$-$19 21 48.0 &- &- \\
      && 496 & 1342205490   &    &              &                         &                  & -1.308&-0.682  \\
 \twob&G11.11-0.12 P2&291 &  1342191498   &  \ntwohp &     & 18 10 28.40 &$-$19 22 29.0 & - &- \\
       && 520 &1342206660  &   &              &                          &                  &-1.298&-0.702  \\

\hline
\end{tabular}
\tablefoot{
\tablefoottext{a}{Name of \ammonia\ source given in \citet{pillai2006b}.}
\tablefoottext{b}{Operational day of the Herschel mission}
\tablefoottext{c}{Herschel identification number}
\tablefoottext{d}{Spur interferes with the listed transition}
\tablefoottext{e}{Requested pointing}
\tablefoottext{f}{Pointing offset (arcsec) based on reconstructed pointing between ODs 320 and 750}
}
\end{center}
\end{table*}

\subsection{The APEX observations}
Complementary observations were carried out between October 2005 and  August 2006 with the 12-m
Atacama Pathfinder Experiment telescope 
\citep[APEX,][]{guesten2006}\footnote{This
  publication is based on data acquired with the Atacama Pathfinder
  Experiment (APEX). APEX is a collaboration between the
  Max-Planck-Institut f\"ur Radioastronomie, the European Southern
  Observatory, and the Onsala Space Observatory.} using the APEX-2 345~GHz facility receiver
\citep{risacher2006}. The receiver was tuned to 279.5~(\ntwohp\  3--2) and 337.8~(\csto\ 3--2) GHz. 
As backend the 2$\times$1~GHz facility FFTS spectrometer
\citep{klein2006} was used and the observations included  the C$^{34}$S 7--6 and CH$_3$OH 7$_K$--6$_K$ lines
along with
 C$^{17}$O 3--2.  To increase the
signal-to-noise the resulting
spectra were smoothed to a velocity resolution of 0.87 \kms. The
observations were carried out under excellent weather conditions
with PWV of 0.5~mm, leading to DSB system temperatures of about
130~K. The focus was adjusted on Jupiter and the pointing was corrected
with continuum cross-scans on the nearby hot core G10.47+0.03.
A beam efficiency of 0.73 was applied to convert the data to the $T_{mb}$ scale.

We note that there is a pointing offset up to 22\arcsec\  between the APEX \ntwohp\ (3--2) observations and the HIFI observations.  Although APEX observation falls within the 38\arcsec\ HIFI beam the utility is questionable and is kept to provide an indication of the line strength assuming the emission is extended.  The offsets are indicated in Table \ref{resultstable}.


\begin{table*}
\begin{center}
\caption{Observed lines \label{observedlines}}
\begin{tabular}{lccccccc}
\hline \hline

 Species                                    & Frequency\tablefootmark{1}    & E$_u$\tablefootmark{2}  &HPBW & Observatory& $\eta_{mb}$ & $T_{sys}$& $\delta v$ \\
                                                 & (GHz)            &   (K)      &   (\arcsec)   &                     &                     &      (K)     &   (\kms)    \\     
\hline                                                                     
o-\hto\ 1$_{10} \to 1_{01}$      & 556.93607   &  61.0   & 38& HIFI 1A& 0.75 & 76 & 0.6\\
o-\htoiso\ 1$_{10} \to 1_{01}$ & 547.67644   &  60.5    & 39& HIFI 1A& 0.75 & 76 & 0.6\\
\ntwohp\ $6 \to 5$                   & 558.96666   &  93.9  & 38 & HIFI 1A& 0.75  &76 & 0.6\\
\ntwohp\ $3 \to 2$                   & 279.511701 &  26.8  & 22 & APEX& 0.73  & 120 -- 210 & 0.53\\
\csto\ $ 3 \to 2 $                     &337.061129  &  32.4  &  18    & APEX  & 0.73 & 130 & 0.87\\
\ctfs\ $7 \to 6$                       &337.396459  &  64.8  &  18    & APEX  & 0.73 & 130 &0.87\\
\cch\ $4_{55} \to 3_{44}$ \tablefootmark{a}	&349.337706&42.0&18& APEX   &0.73 &130 & 0.87\\
\cch\ $4_{44} \to 3_{33}$ \tablefootmark{b}	&349.399276&42.0&18&APEX& 0.73& 130 & 0.87\\
\meth\ $4_0 \to 3_{-1} -$E         &350.687662  &28.0      &  18   & APEX   &0.73 &130 & 0.87\\
\meth\ $7_{-1} \to 6_{-1} -$E      &338.344588  &70.6&  18    &APEX   &0.73&130 & 0.87\\
\meth\ $7_{0 +} \to 6_{0 +} -$A  &338.408698  &65.0&   18    &APEX   &0.73&130 & 0.87\\
\meth\ $7_{0} \to 6_{0} -$E          &338.124488  &78.1&   18    &APEX  &0.73&130 & 0.87 \\
\meth\ $7_{1} \to 6_{1} -$E          &338.614936  &86.1&   18    &APEX  &0.73&130  & 0.87\\
 \meth\ $7_{2} \to 6_{2} -$E         &338.721693  &87.3&   18    &APEX  &0.73&130 &  0.87\\
\meth\ $7_{-2} \to 6_{-2} -$E       &338.722898 &90.9 &   18    &APEX  &0.73&130 & 0.87\\
\meth\ $7_{-2 -} \to 6_{-2 -} -$A&338.512853 &102.7 &  18    &APEX  &0.73&130 & 0.87\\

\hline
\end{tabular}
\tablefoot{References. (1) \citet[CDMS][]{cdms2005}; (2) \citet[LAMDA][]{schoeier2005}
\tablefoottext{a}{Line possibly blended with $4_{5 4} - 3_{4 3}$ at 349.338988 GHz}
\tablefoottext{b}{Line possibly blended with $4_{4 3} - 3_{3 2}$ at 349.400671 GHz}
}
\end{center}
\end{table*}

\section{Results \label{results}}
\begin{figure*}
\begin{center}
\begin{minipage}{0.245\textwidth}
\centering
\includegraphics[width=3.5cm]{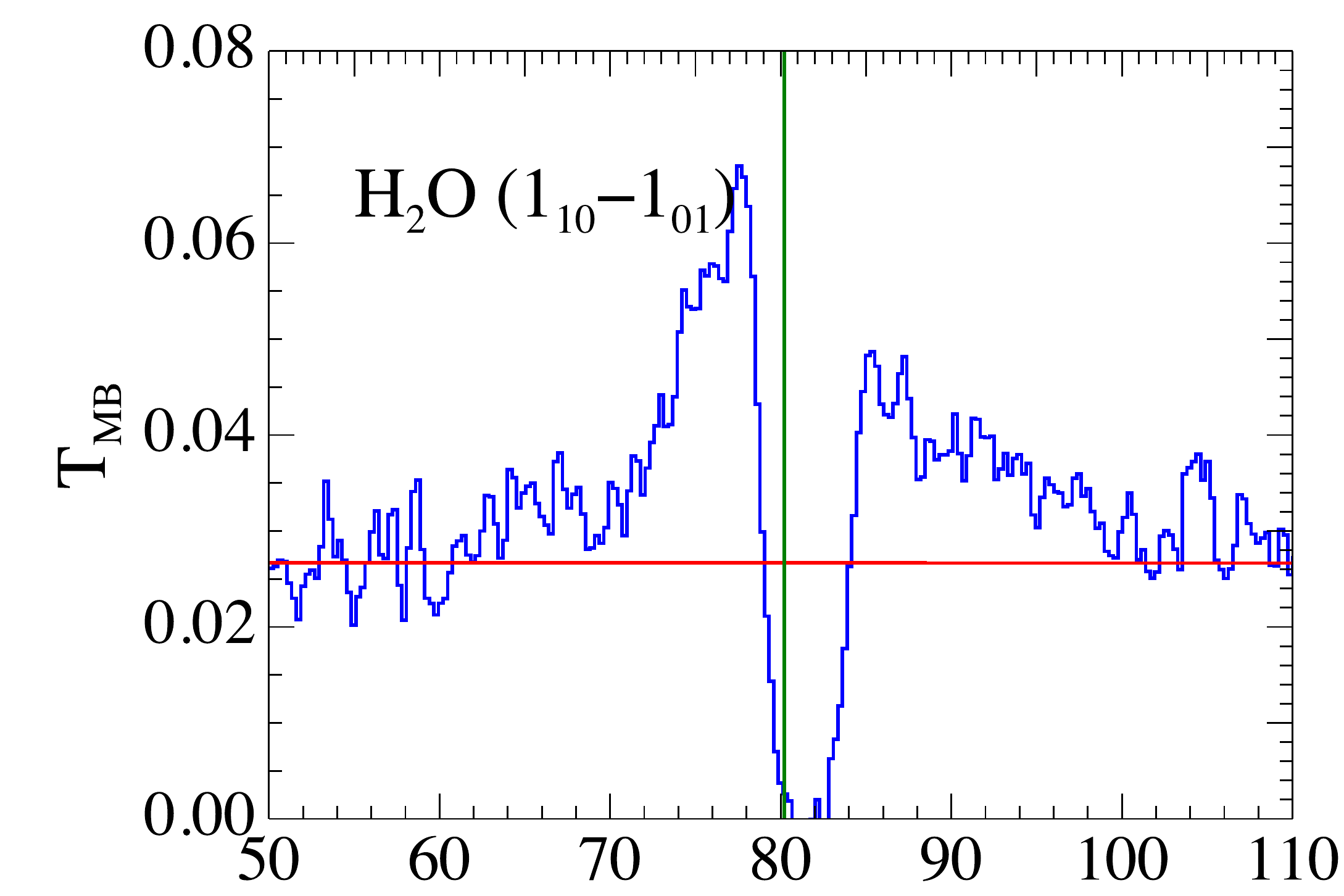}
\end{minipage}
\begin{minipage}{0.245\textwidth}
\centering
\includegraphics[width=3.5cm]{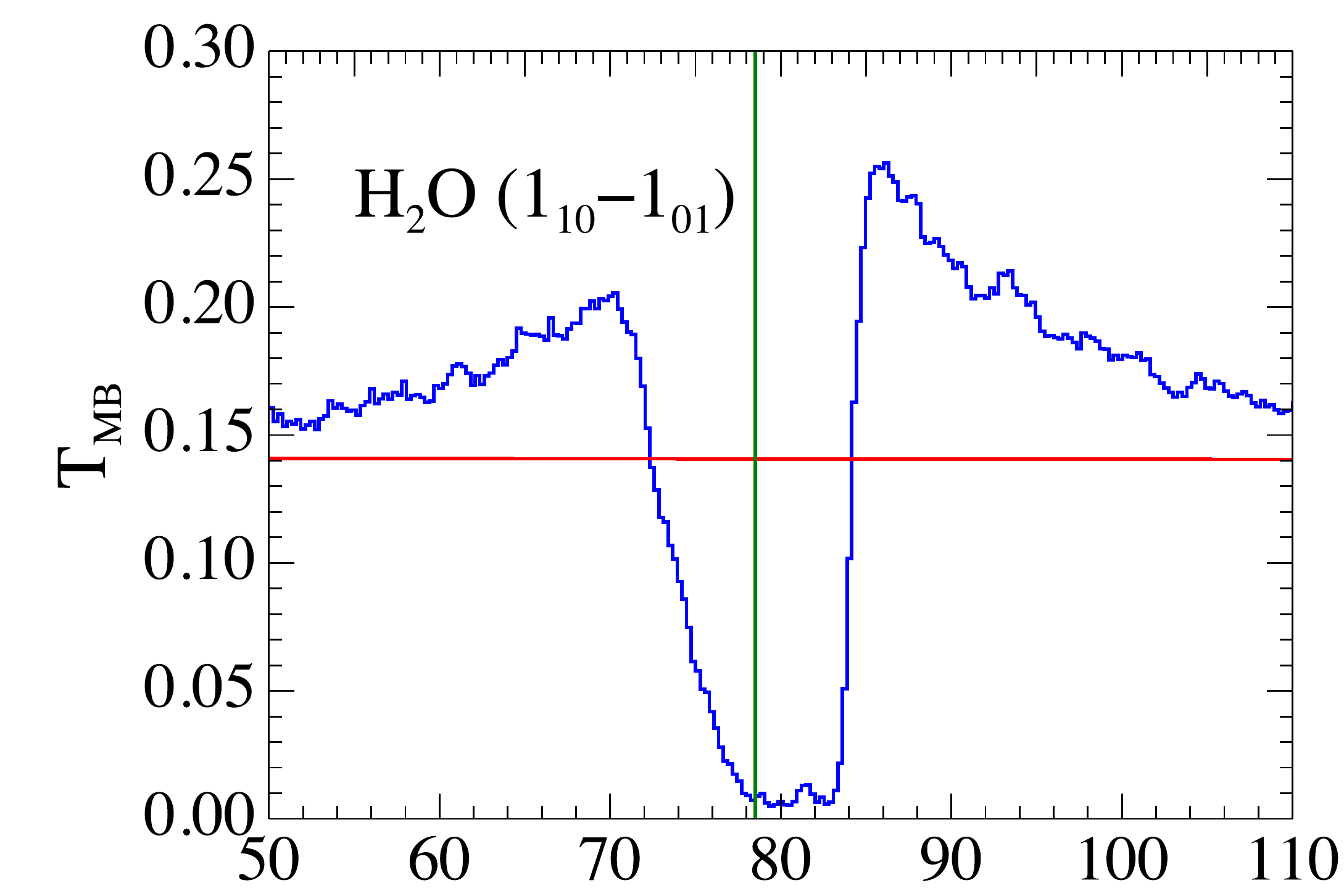}
\end{minipage}
\begin{minipage}{0.245\textwidth}
\centering
\includegraphics[width=3.5cm]{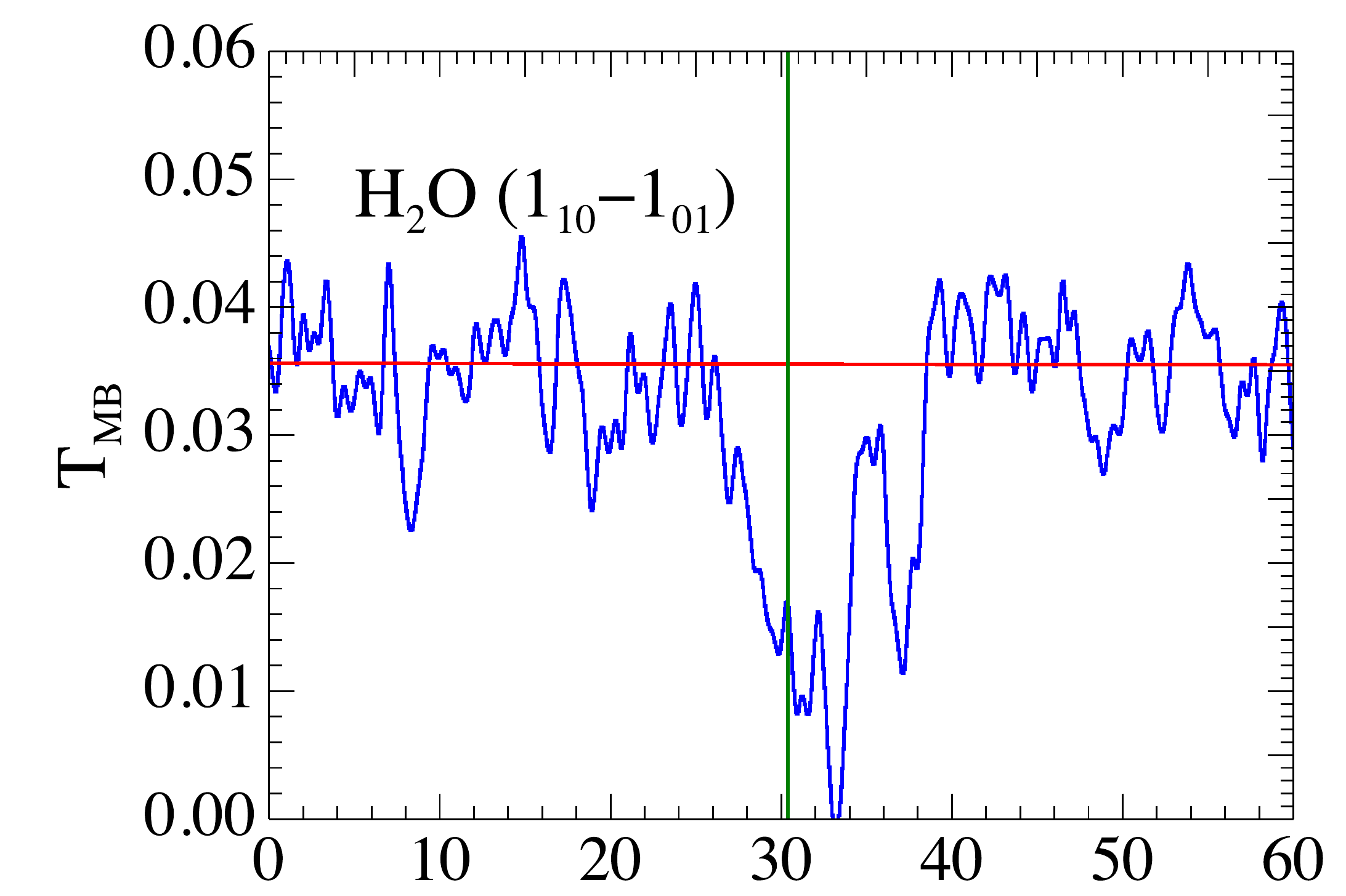}
\end{minipage}
\begin{minipage}{0.245\textwidth}
\centering
\includegraphics[width=3.5cm]{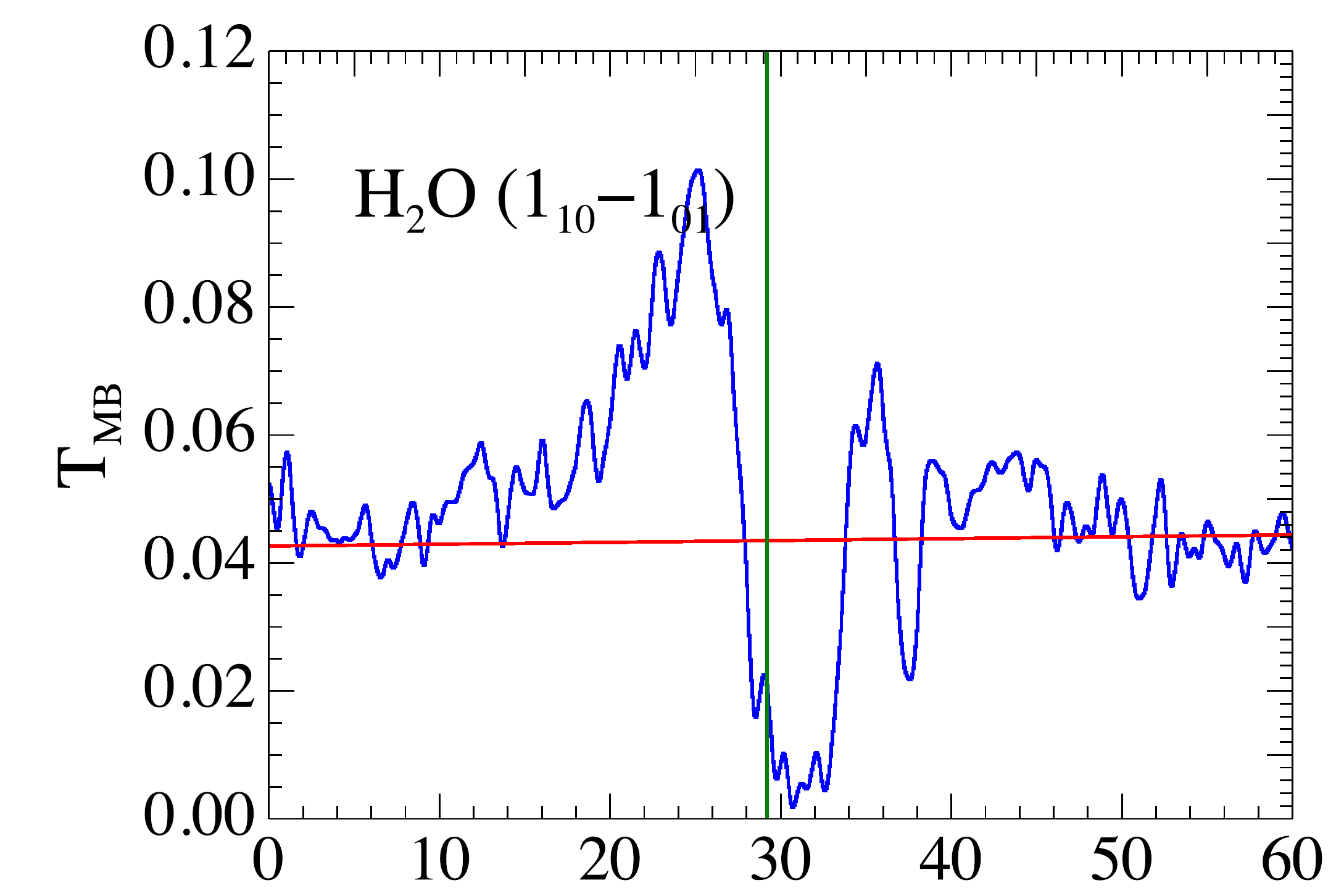}
\end{minipage}
\newpage
\begin{minipage}{0.245\textwidth}
\centering
\includegraphics[width=3.5cm]{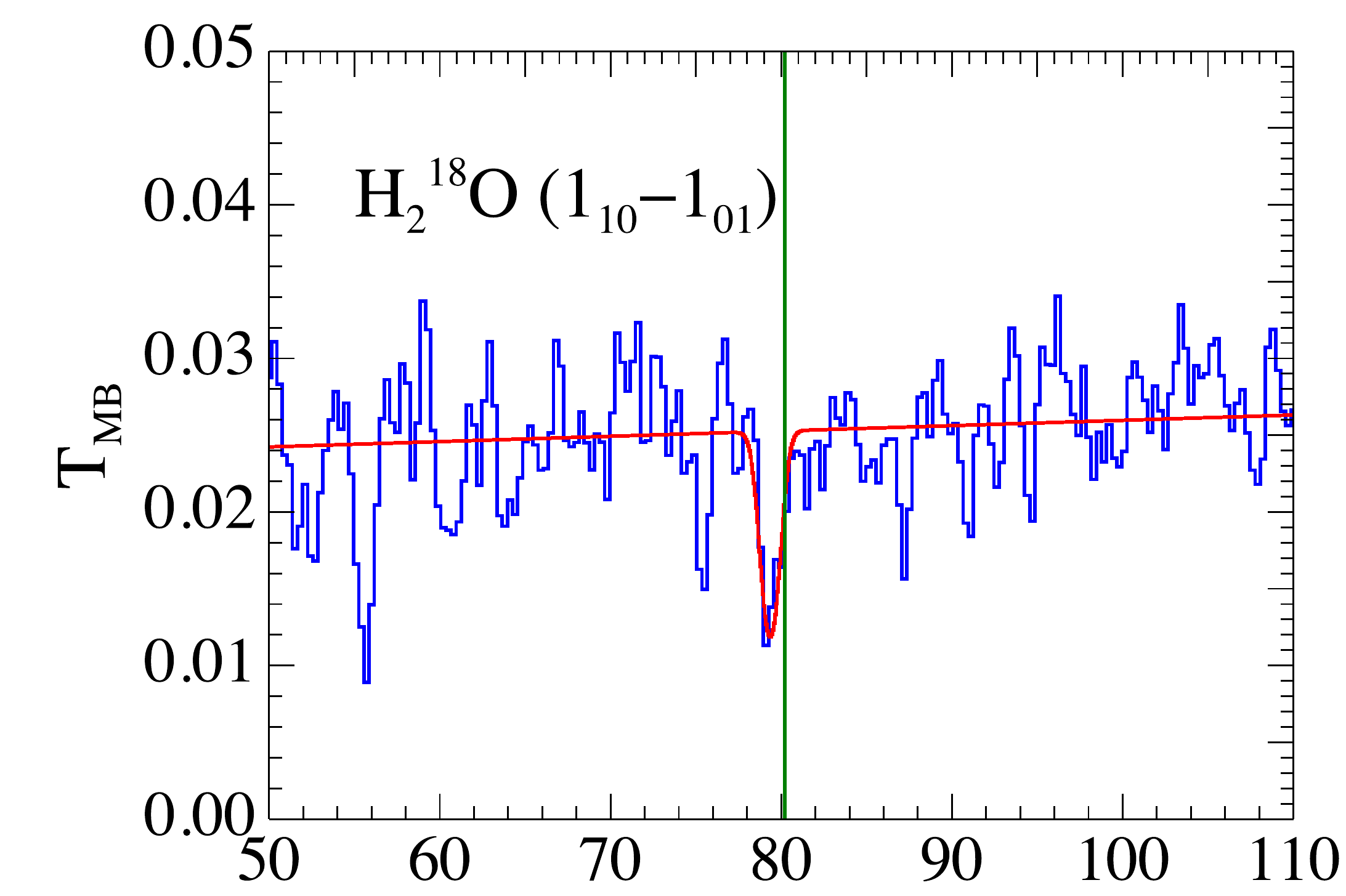}
\end{minipage}
\begin{minipage}{0.245\textwidth}
\centering
\includegraphics[width=3.5cm]{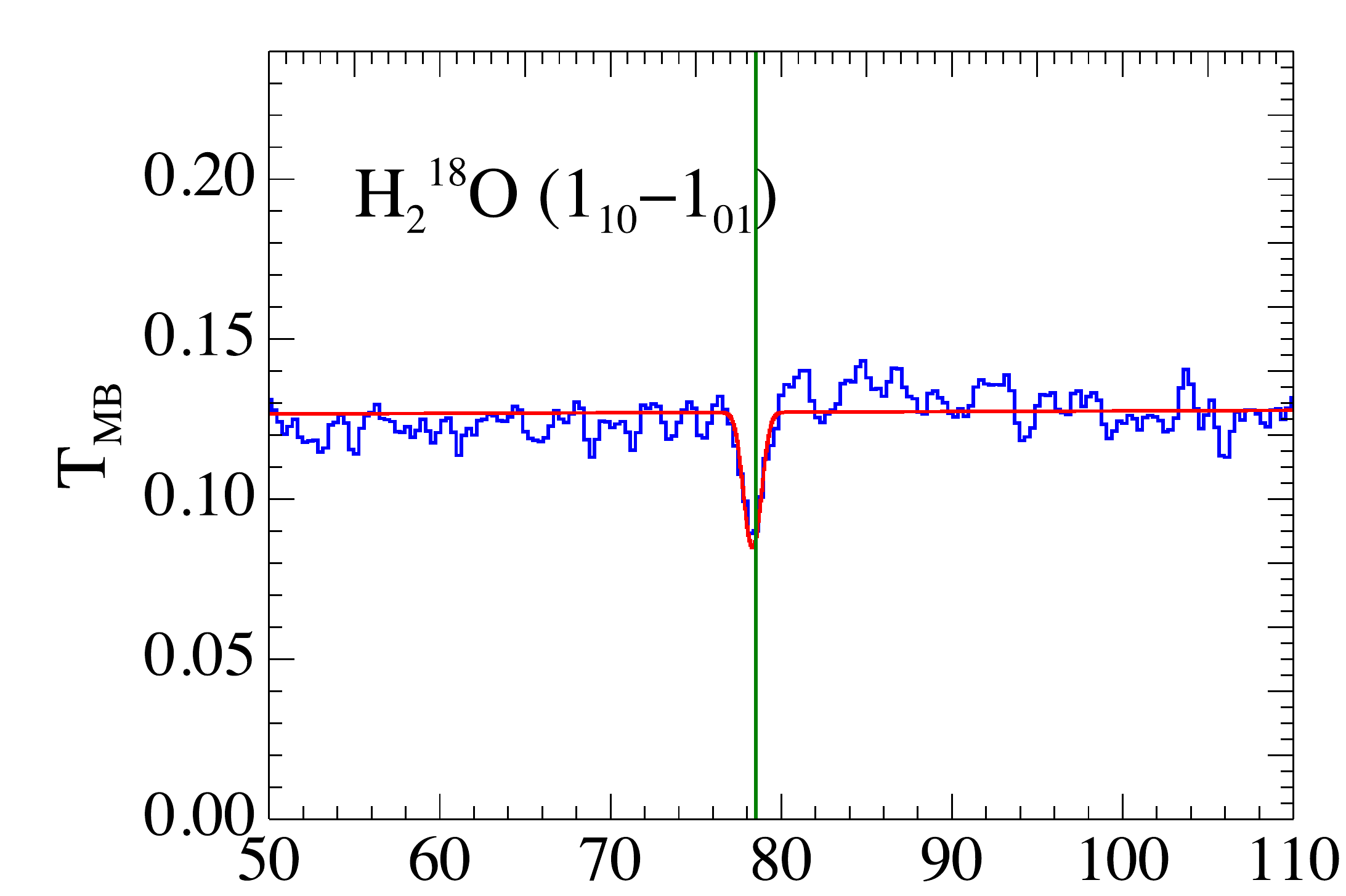}
\end{minipage}
\begin{minipage}{0.245\textwidth}
\centering
\includegraphics[width=3.5cm]{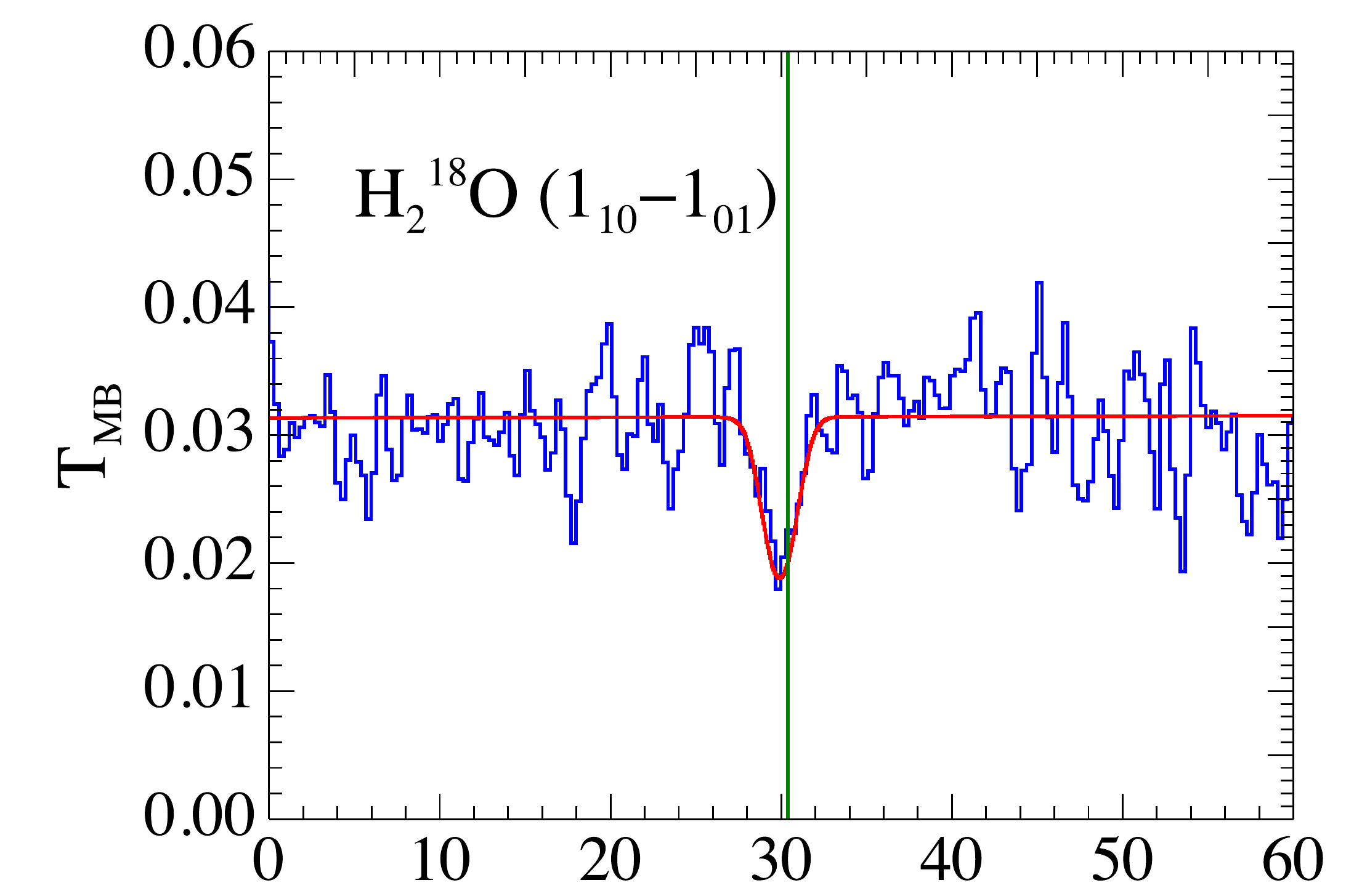}
\end{minipage}
\begin{minipage}{0.245\textwidth}
\centering
\includegraphics[width=3.5cm]{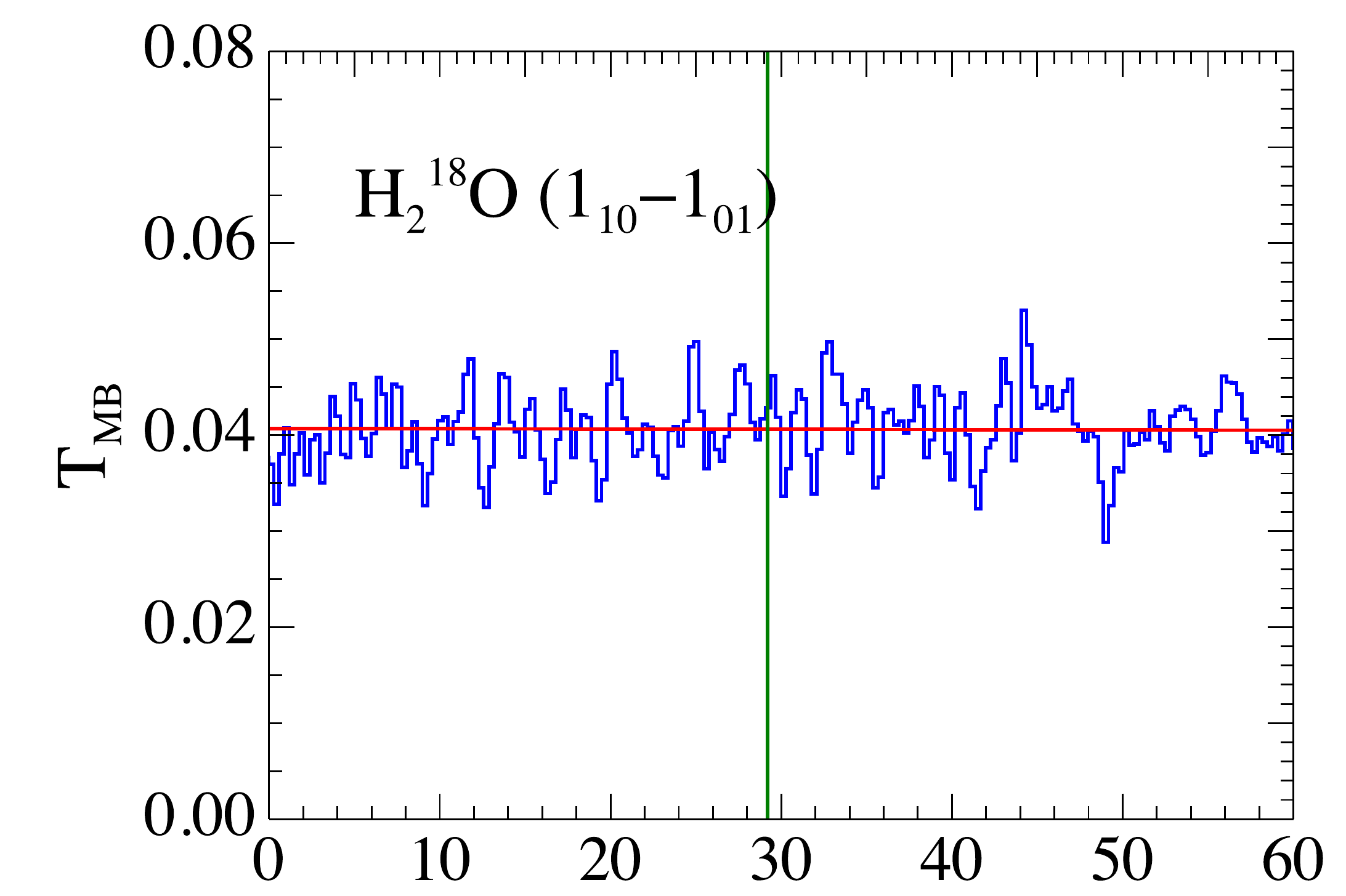}
\end{minipage}
\newpage
\begin{minipage}{0.245\textwidth}
\centering
\includegraphics[width=3.5cm]{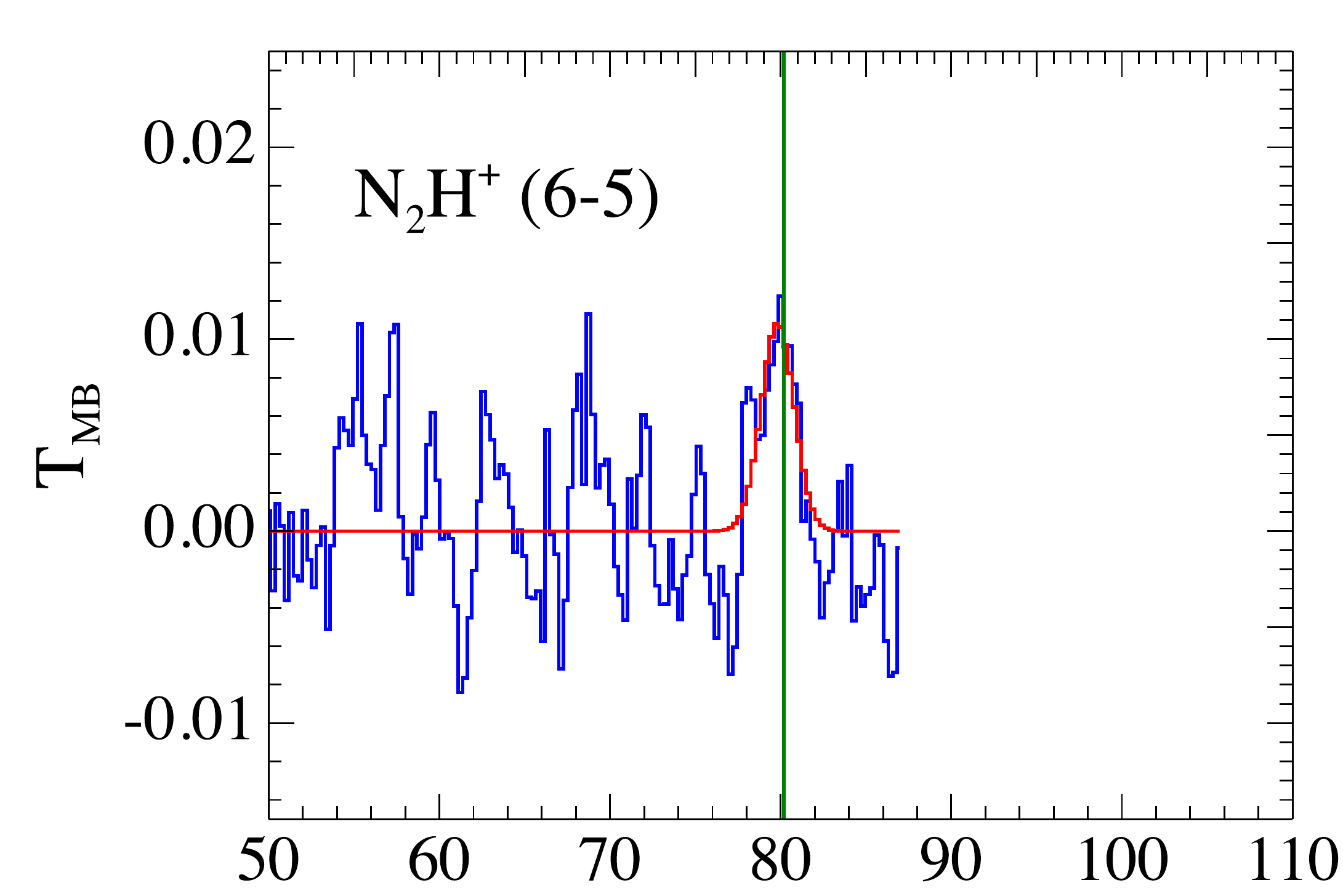}
\end{minipage}
\begin{minipage}{0.245\textwidth}
\centering
\includegraphics[width=3.5cm]{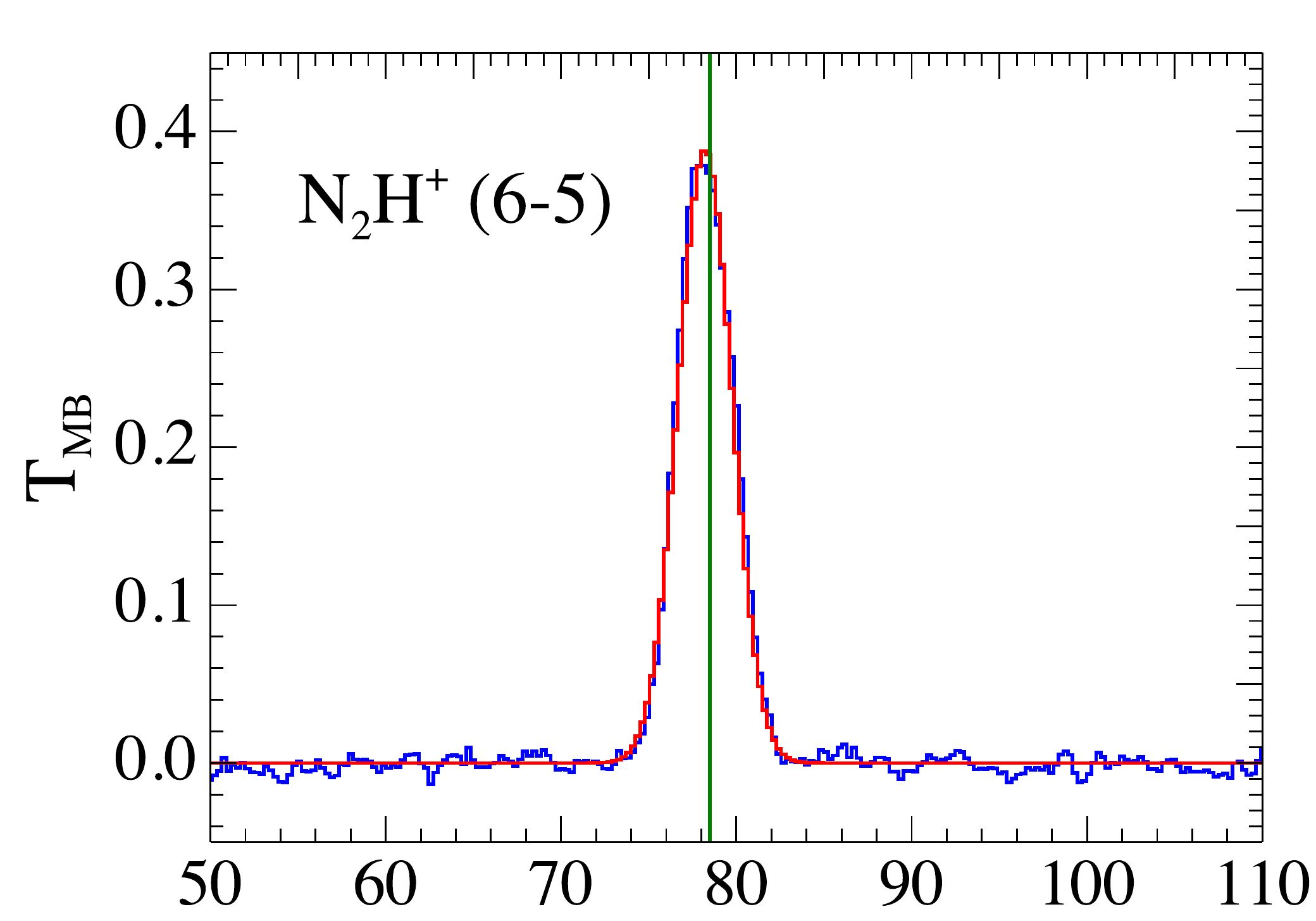}
\end{minipage}
\begin{minipage}{0.245\textwidth}
\centering
\includegraphics[width=3.5cm]{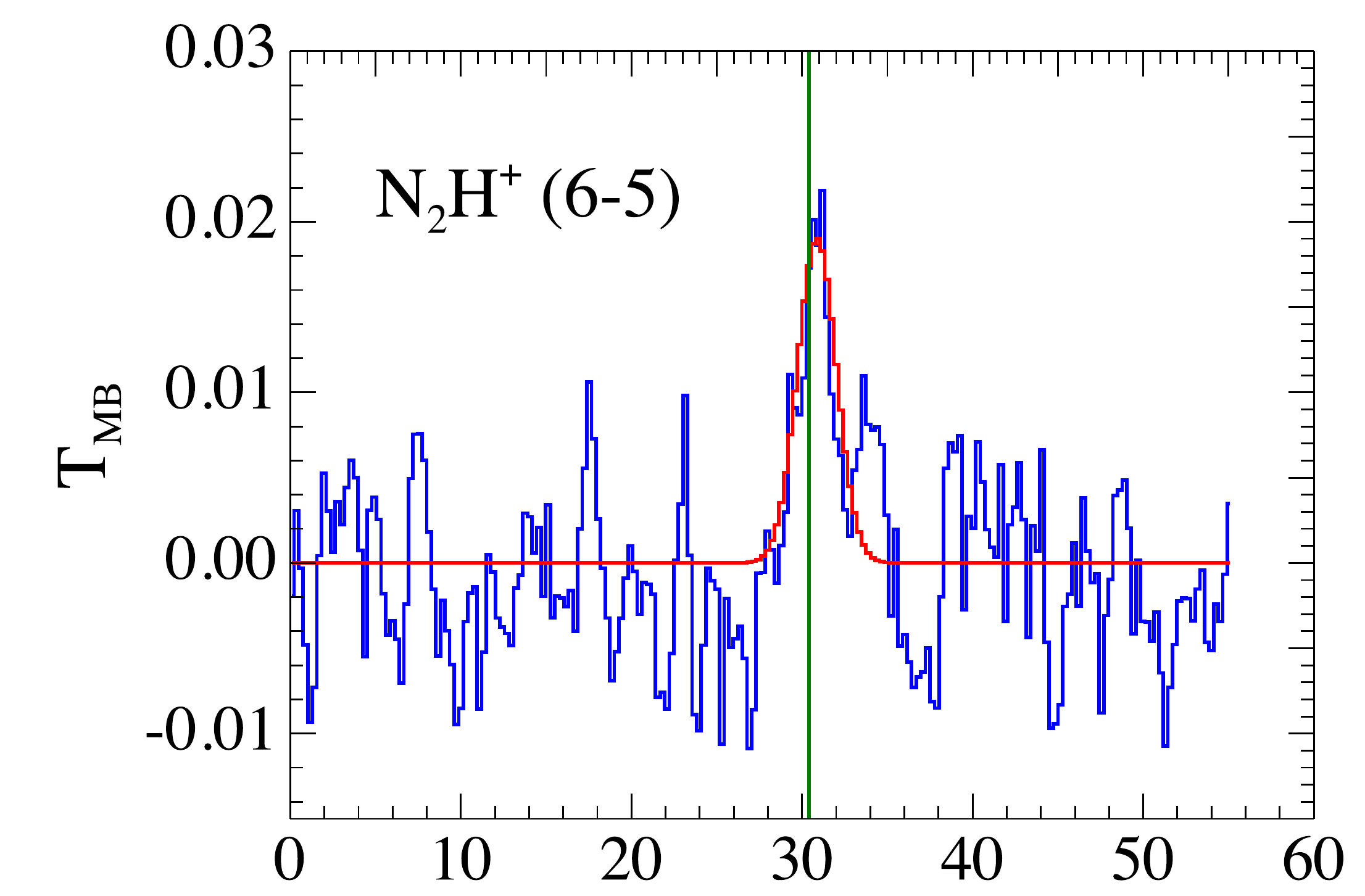}
\end{minipage}
\begin{minipage}{0.245\textwidth}
\centering
\includegraphics[width=3.5cm]{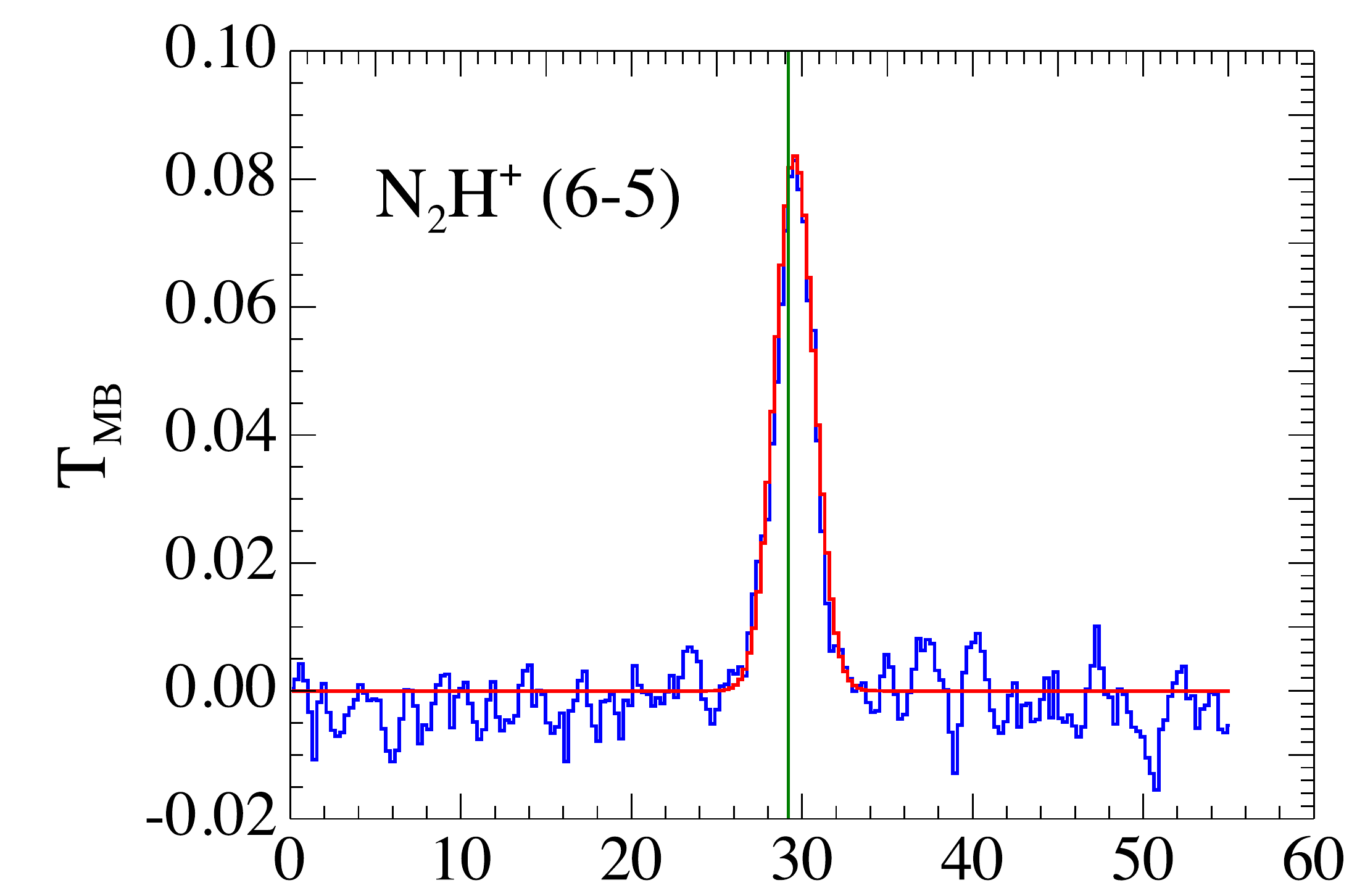}
\end{minipage}
\newpage

\begin{minipage}{0.245\textwidth}
\centering
\includegraphics[width=3.5cm]{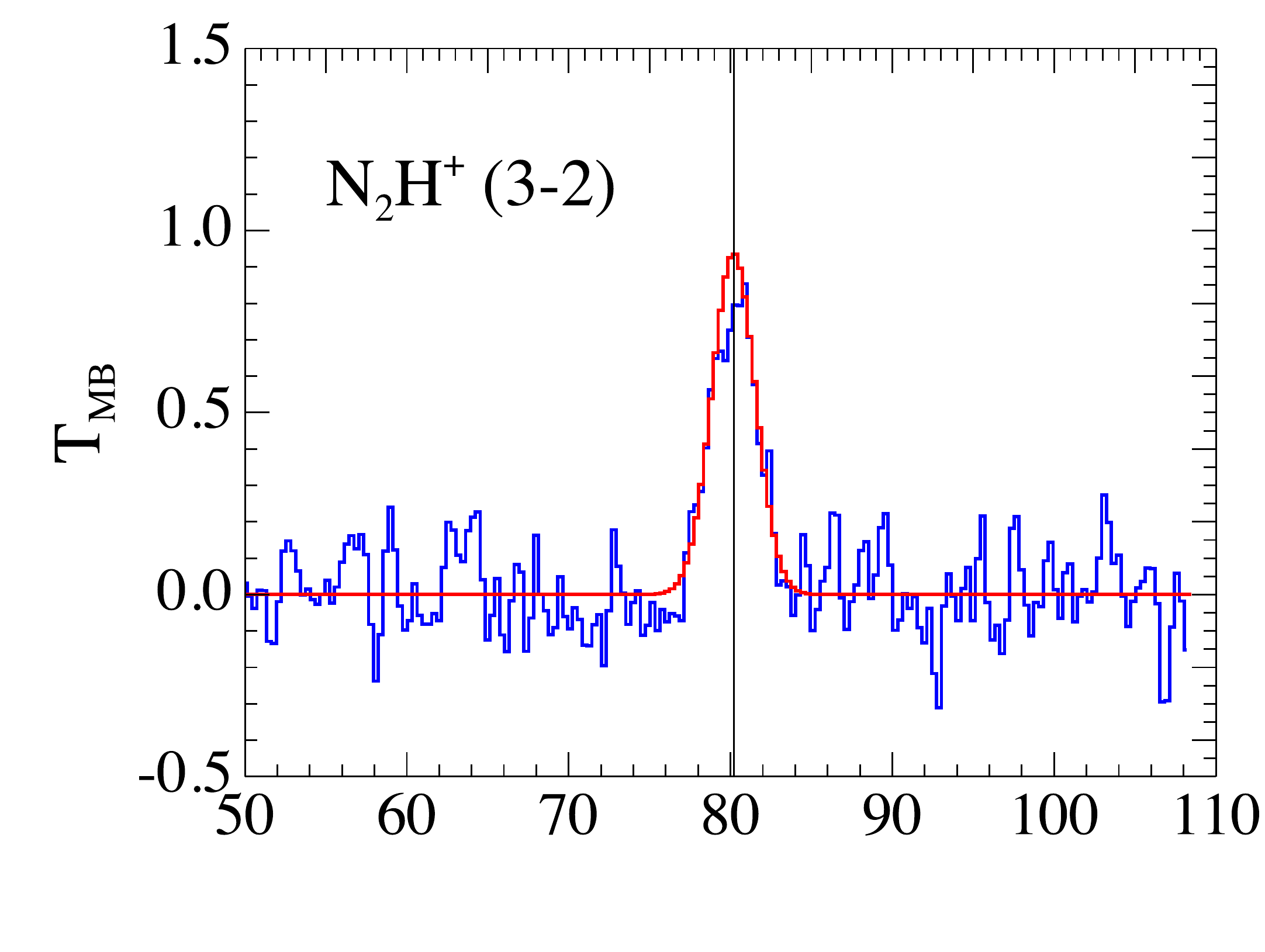}
\end{minipage}
\begin{minipage}{0.245\textwidth}
\centering
\includegraphics[width=3.5cm]{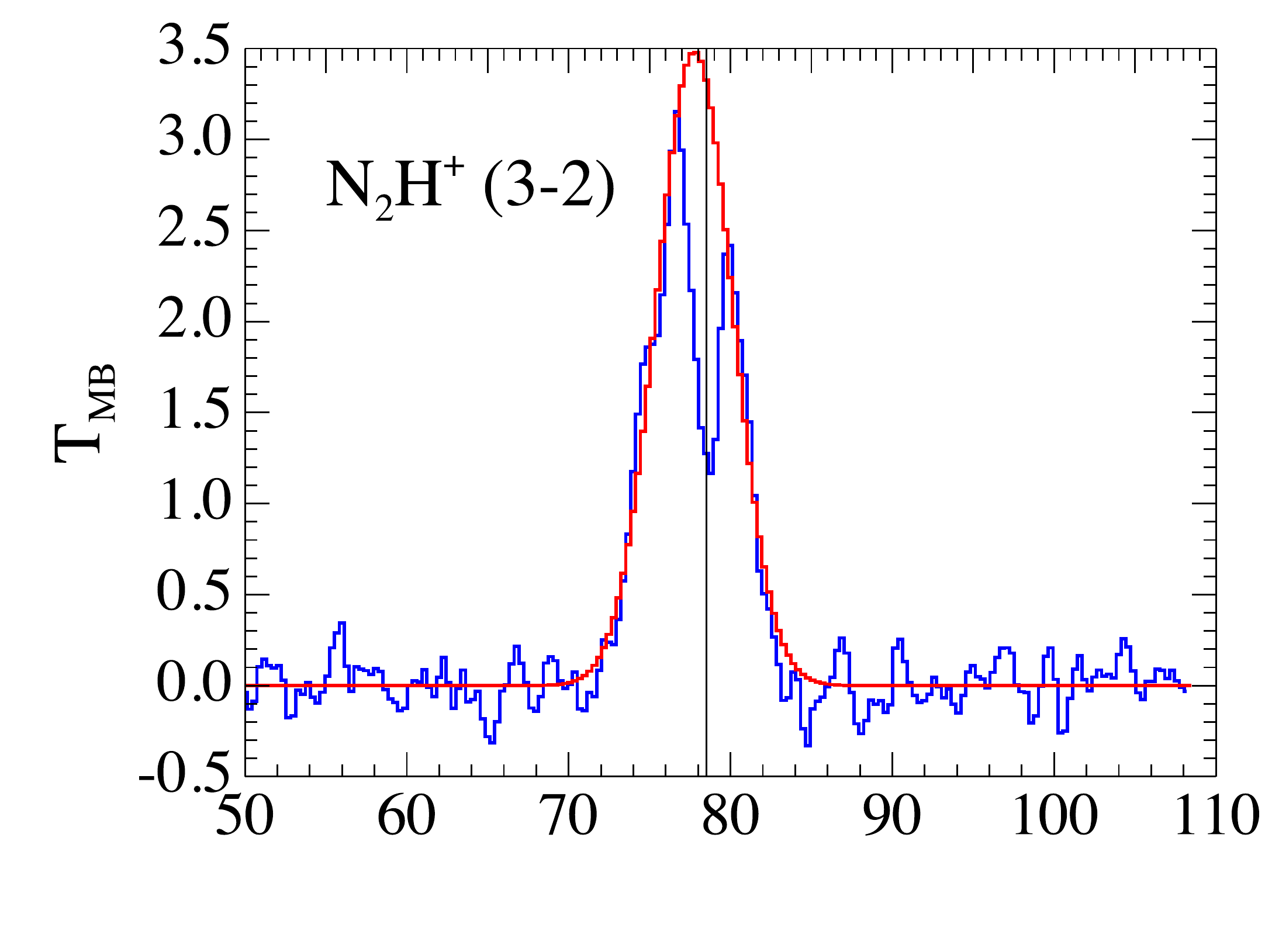}
\end{minipage}
\begin{minipage}{0.245\textwidth}
\centering
\includegraphics[width=3.5cm]{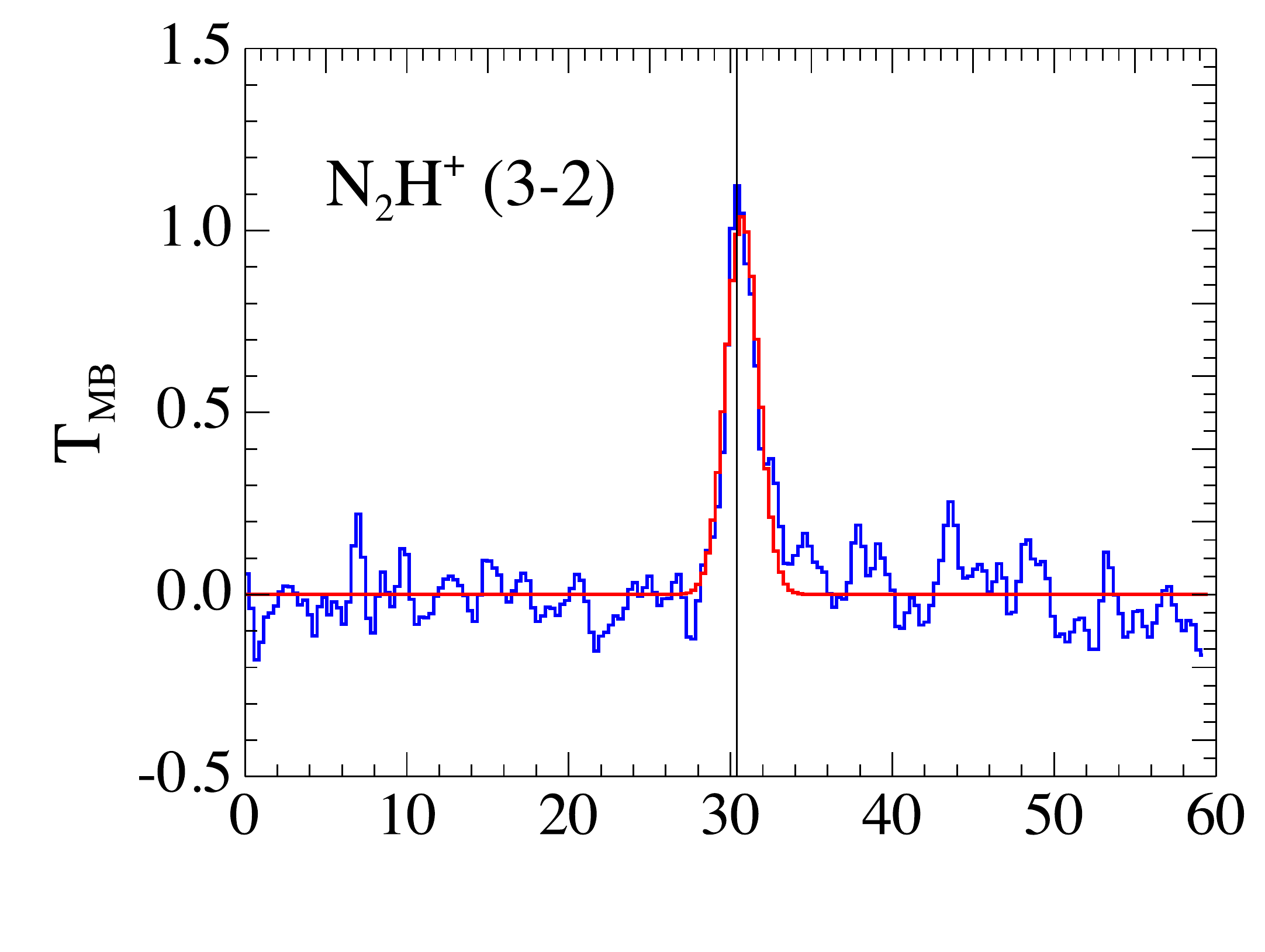}
\end{minipage}
\begin{minipage}{0.245\textwidth}
\centering
\includegraphics[width=3.5cm]{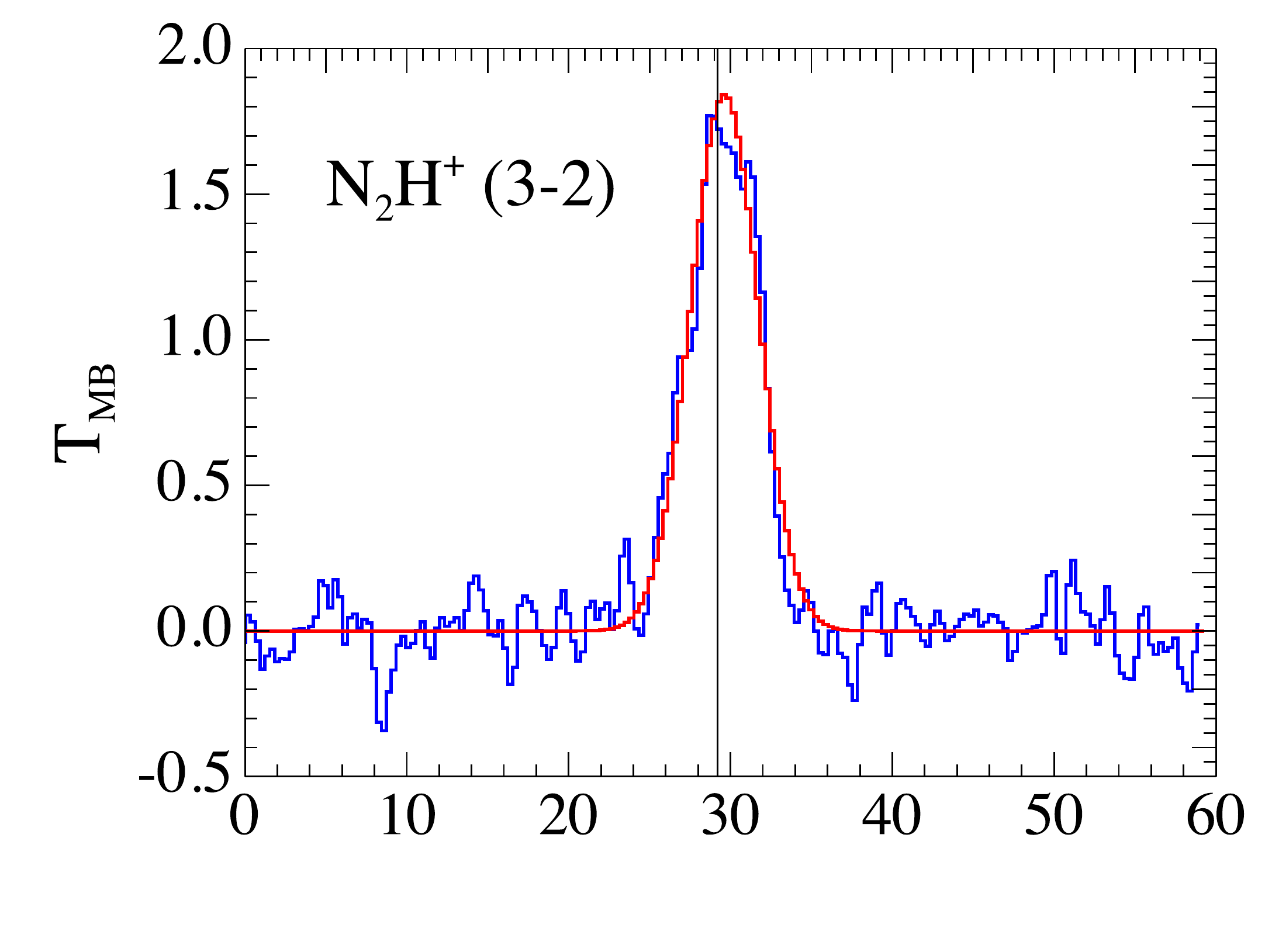}
\end{minipage}
\newpage

\begin{minipage}{0.245\textwidth}
\centering
\includegraphics[width=3.5cm]{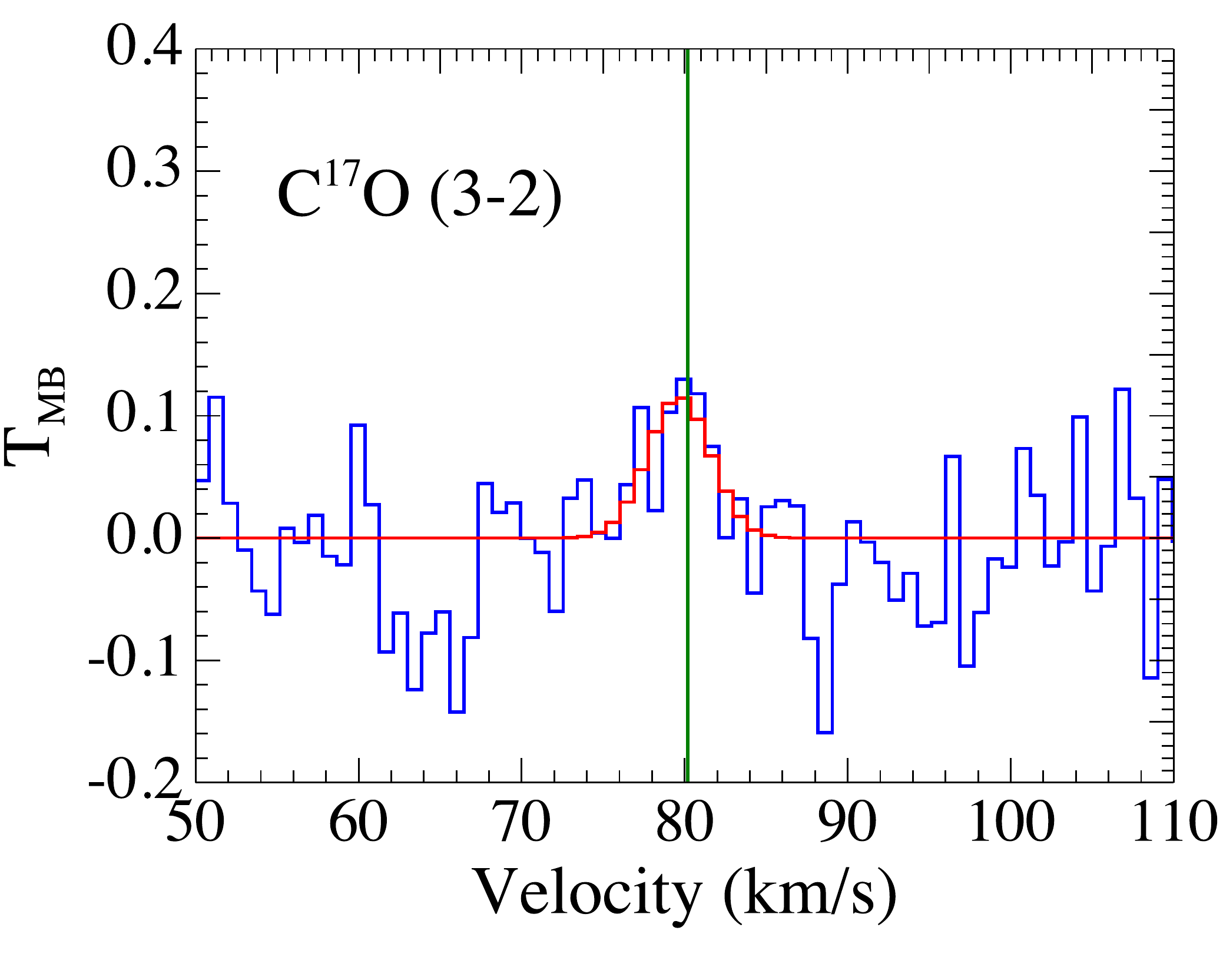}

\onea
\end{minipage}
\begin{minipage}{0.245\textwidth}
\centering
\includegraphics[width=3.5cm]{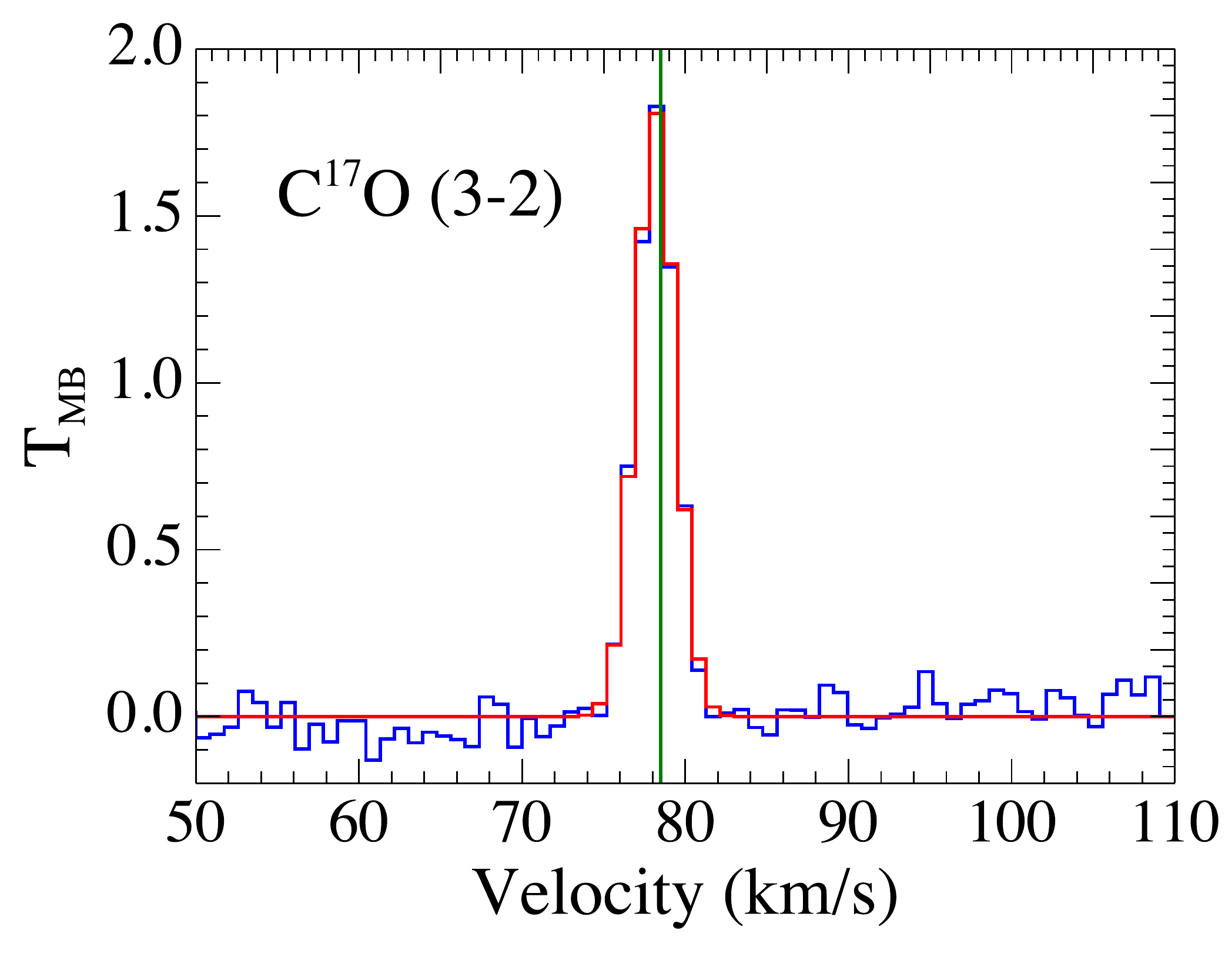}
\oneb
\end{minipage}
\begin{minipage}{0.245\textwidth}
\centering
\includegraphics[width=3.5cm]{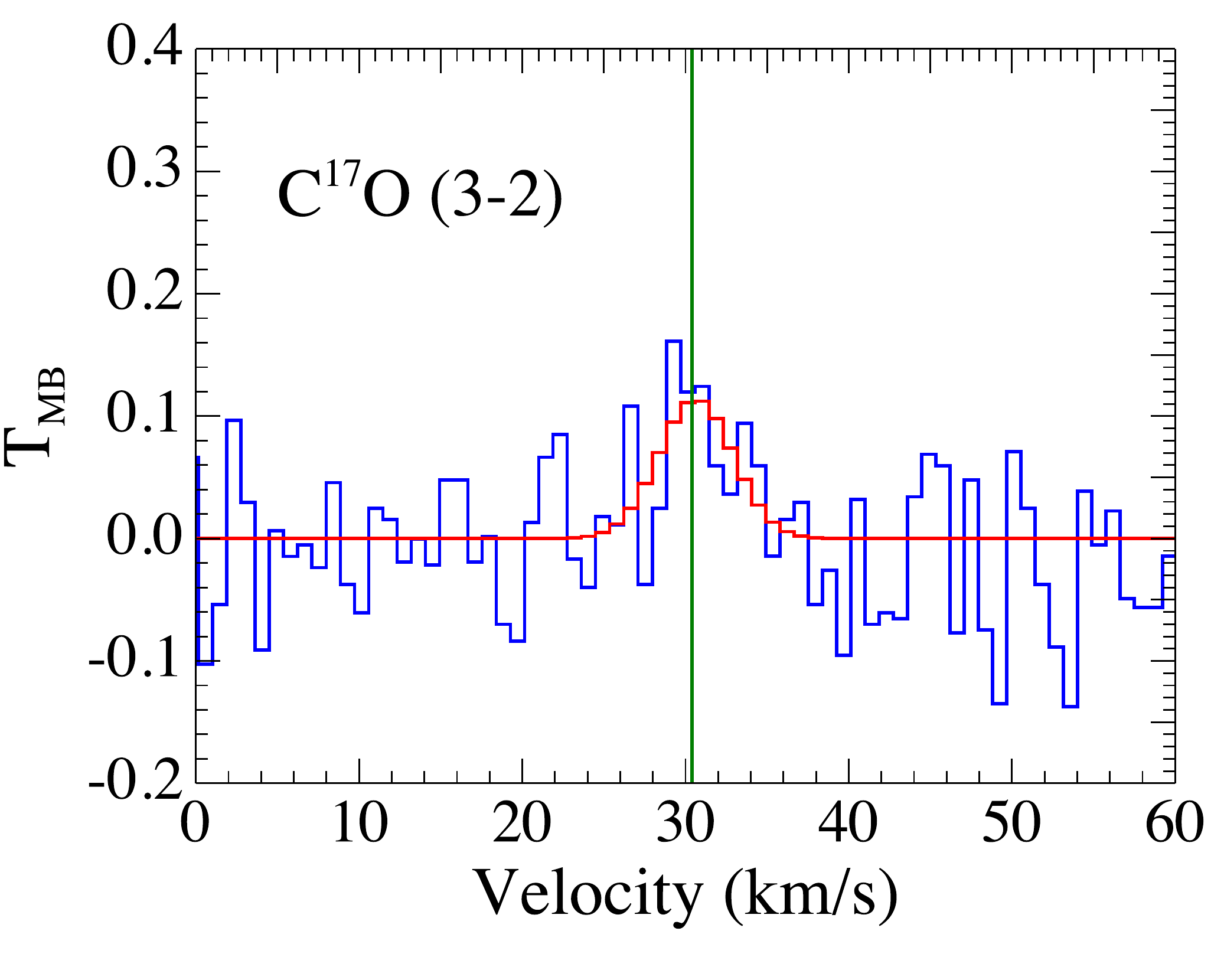}
\twoa
\end{minipage}
\begin{minipage}{0.245\textwidth}
\centering
\includegraphics[width=3.5cm]{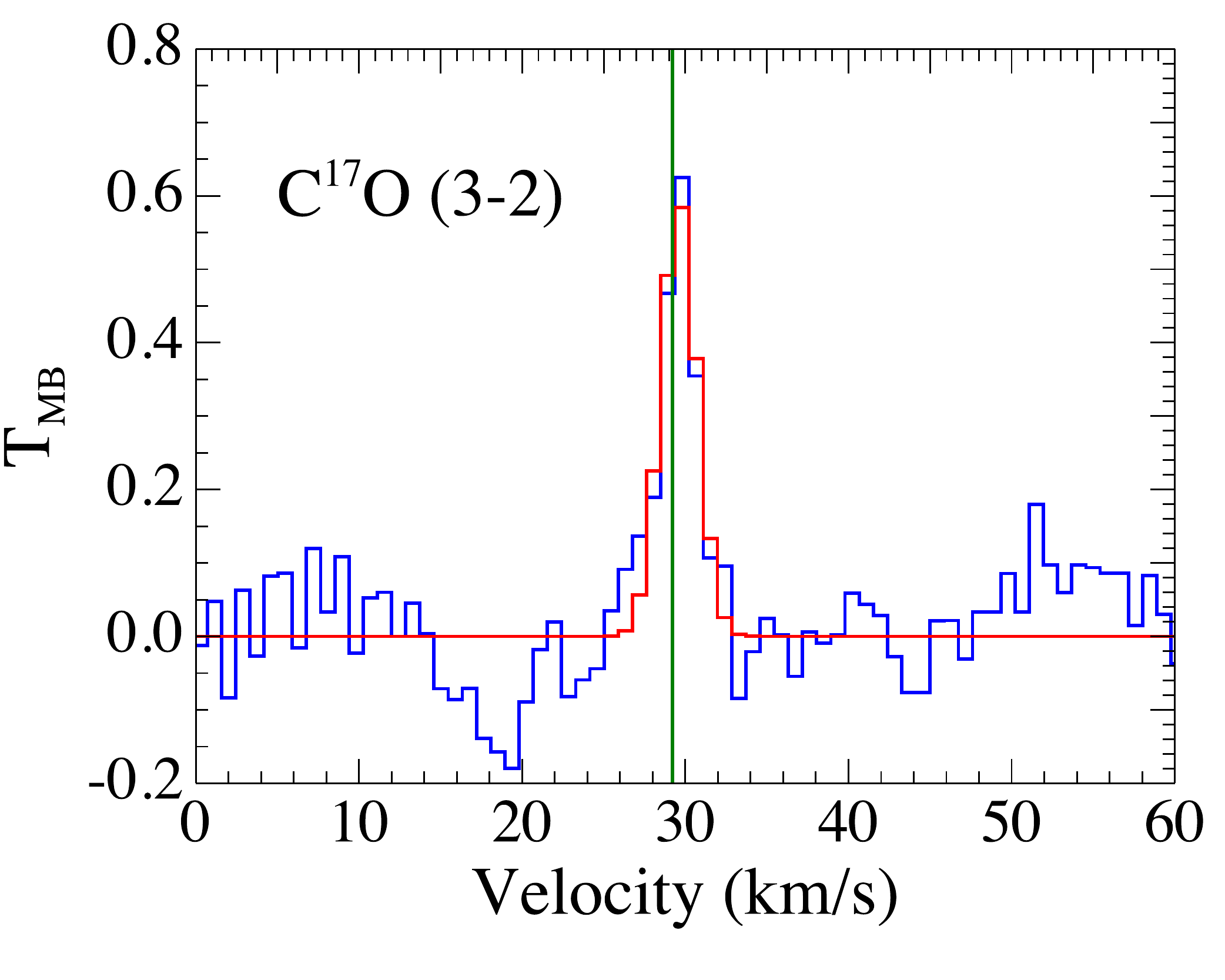}
\twob
\end{minipage}
\caption{Data for HIFI  and APEX (\csto\ \bf \rm  and \ntwohp \rm) observations for each \bf \rm  clump\rm and transition are shown in blue.  Gaussian fits are indicated in red \bf \rm  except for the water transition where only the continuum level is shown in red\rm.  The systemic velocity of each \bf \rm  clump \rm is indicated with a green line.  The complex profiles of the main \hto\ line are clearly visible.  The full model decompositions of both emission and absorption are shown in Figs. \ref{modelobs} and \ref{modelabs} respectively.  \label{observedlineplots}}
\end{center}
\end{figure*}

Figure \ref{observedlineplots} shows the observations of the main lines from Table \ref{observedlines}.   All four \bf \rm  clumps \rm show saturated \hto\ absorption and three of the four \bf \rm  clumps \rm also have broad emission with self absorption.   The clump \twoa\ only shows absorption.  The \htoiso\ line is seen in three sources the exception being \twob.    The \csto\ (3--2), \ntwohp\ (3--2) and \ntwohp\ (6--5) emission lines are detected at all four positions.  The \ntwohp\ (3--2) data for \oneb\ indicate self absorption.   The \csto\ (3--2) and \ntwohp\ (6--5) data for \onea\ are marginal detections while the \meth\ and \ctfs\ lines (not shown here) are detected in the \submm\ \bf \rm  clumps \rm: \oneb\ and \twob.  

The emission lines are parametrized by fitting single or multiple Gaussians to the data excluding the absorption features.  All the \hto\ emission lines display complex emission lines which are well represented by two broad Gaussian components.  Table \ref{resultstable} lists the results of Gaussian fits to each line. 

The procedure used to describe the \hto\ and \htoiso\ absorption lines is described below.


\subsection{Fitting of  absorption-emission features \label{fittingprocedure}}
\begin{figure}

\includegraphics[width=8.0cm]{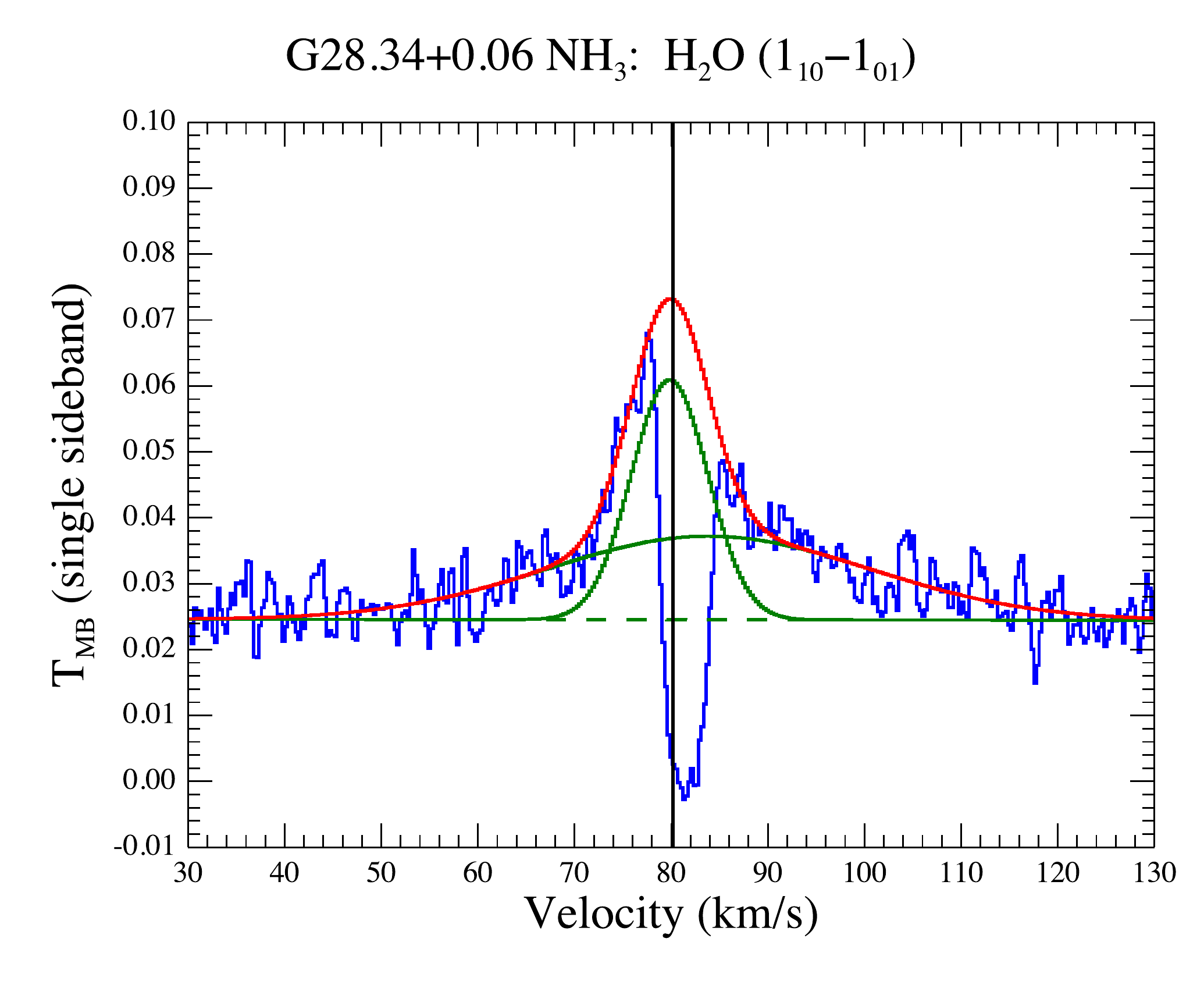}

\includegraphics[width=8.0cm]{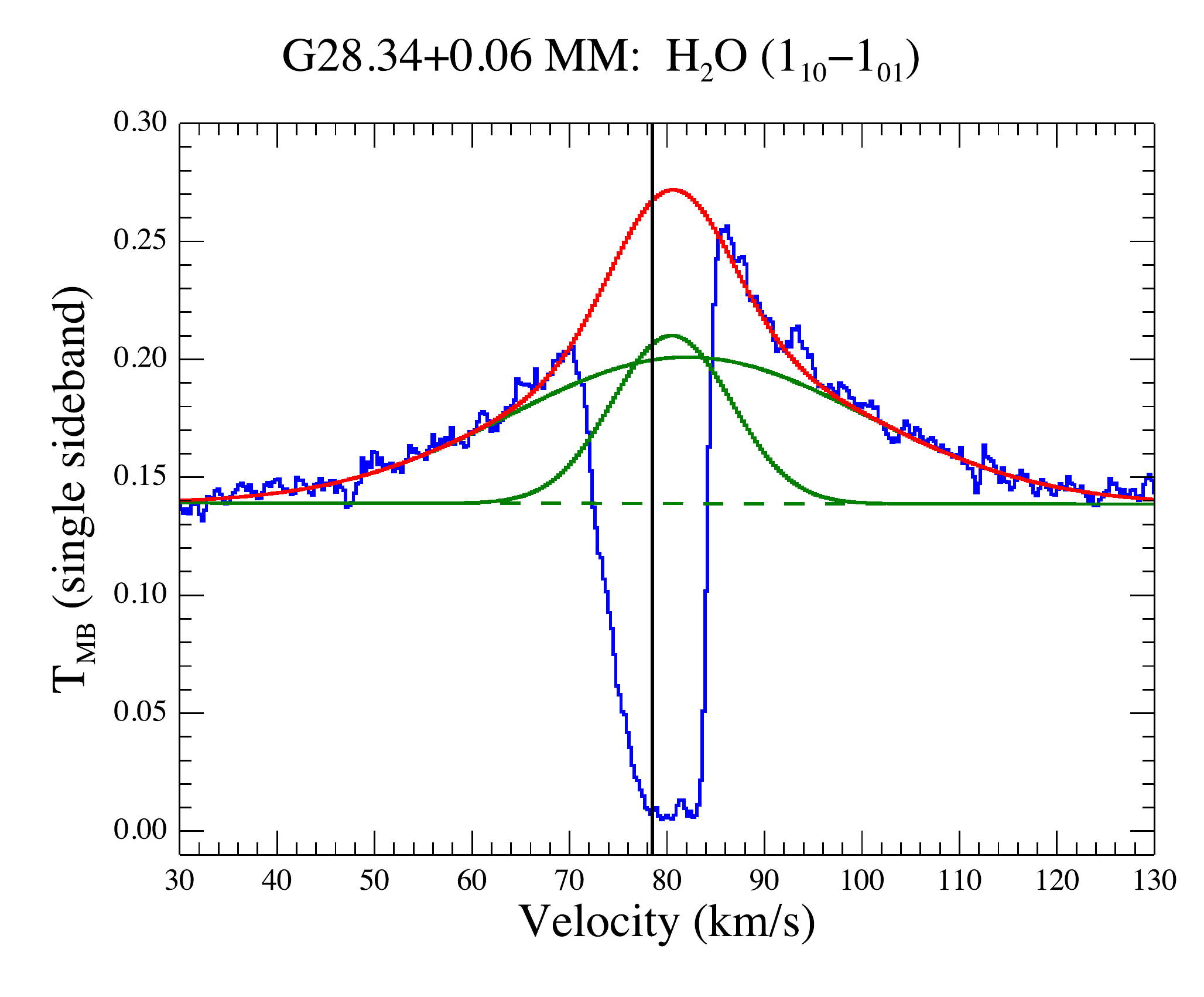}

\includegraphics[width=8.0cm]{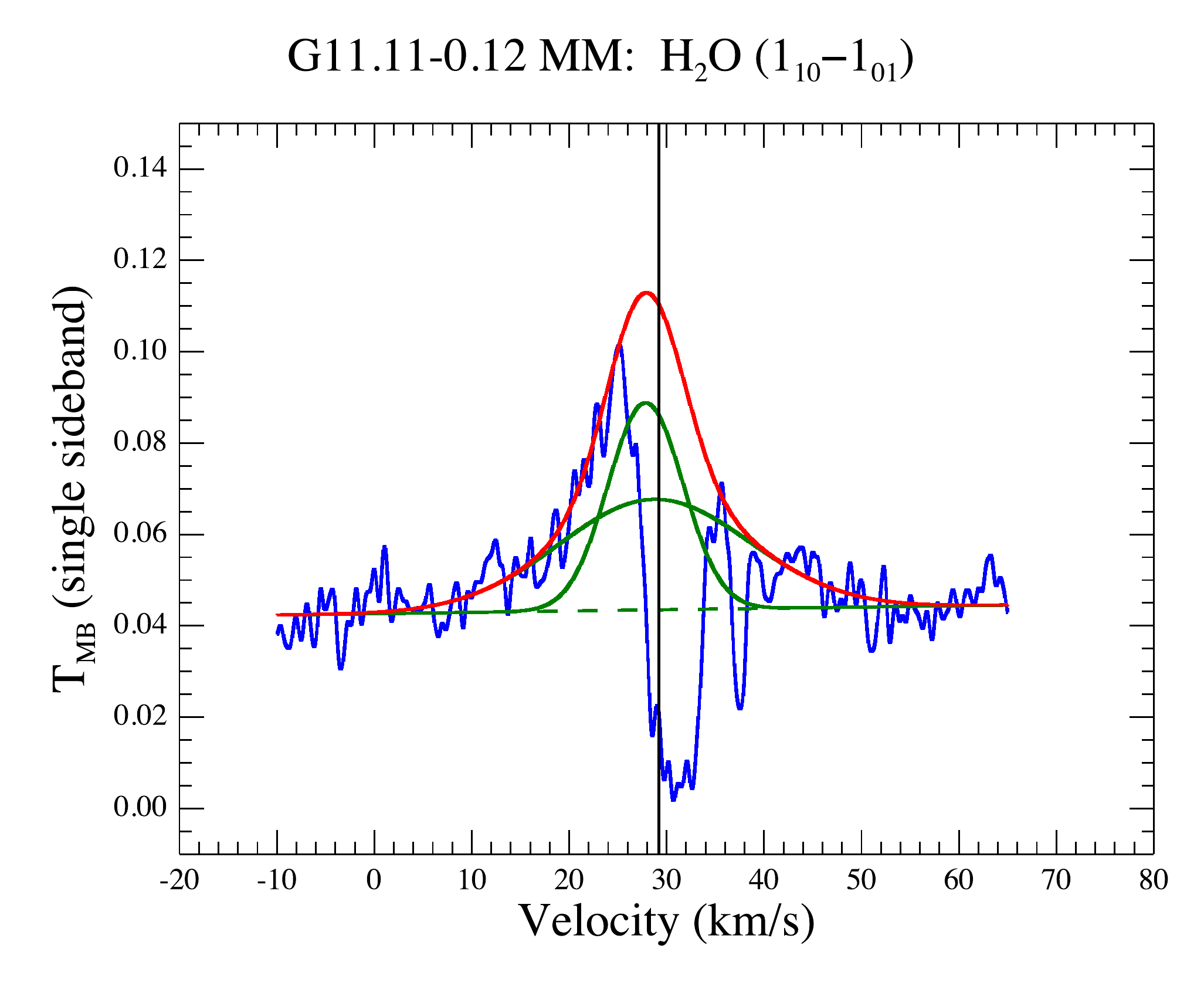}

\caption{Water emission seen in three \bf \rm  clumps \rm \label{modelobs}.  Emission lines are fit with two Gaussian components \bf \rm (solid green).  The continuum is a first order polynomial fit (dashed green). \rm The total emission is the sum of the two components and the continuum  (shown in red). }
\end{figure}

In this work, emission and absorption lines are treated differently.  The emission lines are parametrized with single or multiple Gaussian fits.  The continuum emission is taken as a first order polynomial.  The total emission is the sum of all the components,
\[
I_{\nu} = I_{\nu cont}+\sum_i G_{\nu}(I_{0,i},\nu_{0,i},\Delta \nu_i),
\]
where I$_{\nu cont}$ is the single sideband estimate of the continuum and  G$_{\nu}$ is the Gaussian parametrized by the intensity I$_0$, central frequency $\nu_0$ and width $\Delta \nu_0$ for each emission component.  The data used to determine these parameters are at frequencies which do not show signs of absorption.   The emission lines and continuum are shown in Fig. \ref{modelobs}. The  sum of the resulting Gaussian curves and the continuum then represents the total emission.   

To measure the absorption lines, the data are normalized by dividing out the total emission.   Normalized data are then fit for a Gaussian absorption component.  The total absorption is the product of each absorbing component, 
\[
I_{\nu,norm} = \prod_i exp^{-\tau_{\nu ,i}},
\]
where $\tau_{\nu}$ is itself a Gaussian absorption feature.    

\begin{figure*}
\begin{center}
\begin{minipage}{0.45\textwidth}
\centering
\includegraphics[width=6.0cm]{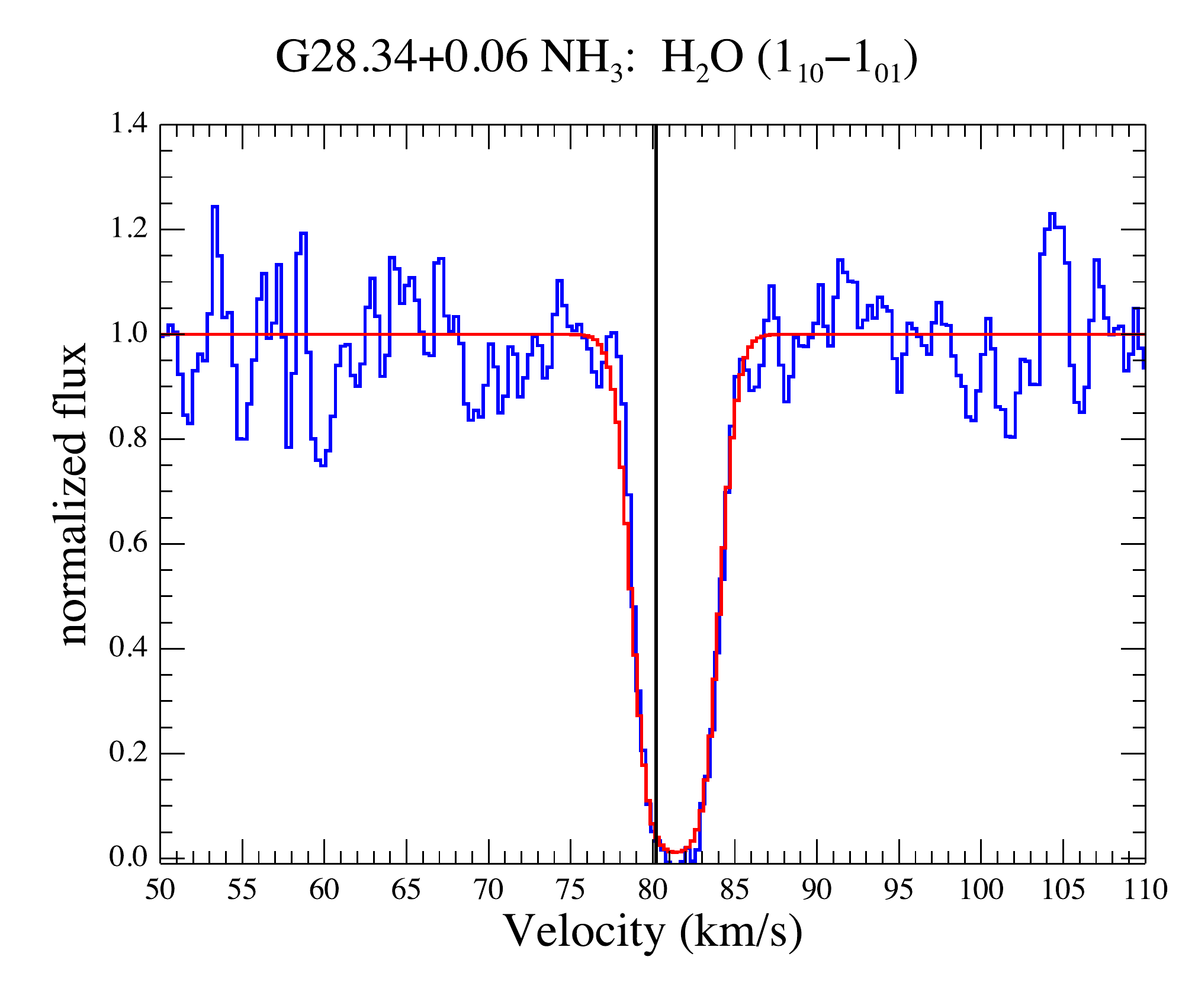}
\end{minipage}
\begin{minipage}{0.45\textwidth}
\centering
\includegraphics[width=6.0cm]{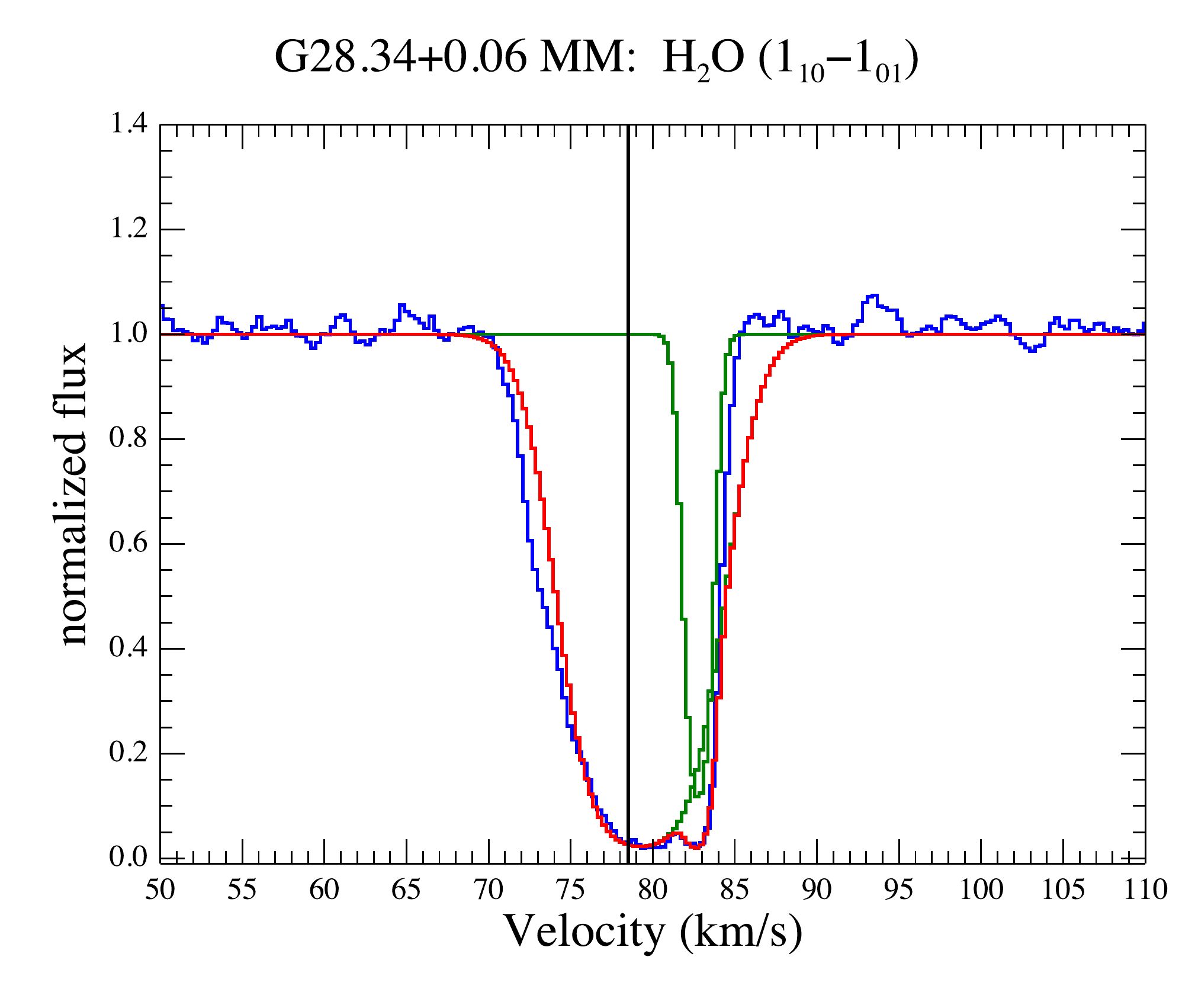}
\end{minipage}
\newpage
\begin{minipage}{0.45\textwidth}
\centering
\includegraphics[width=6.0cm]{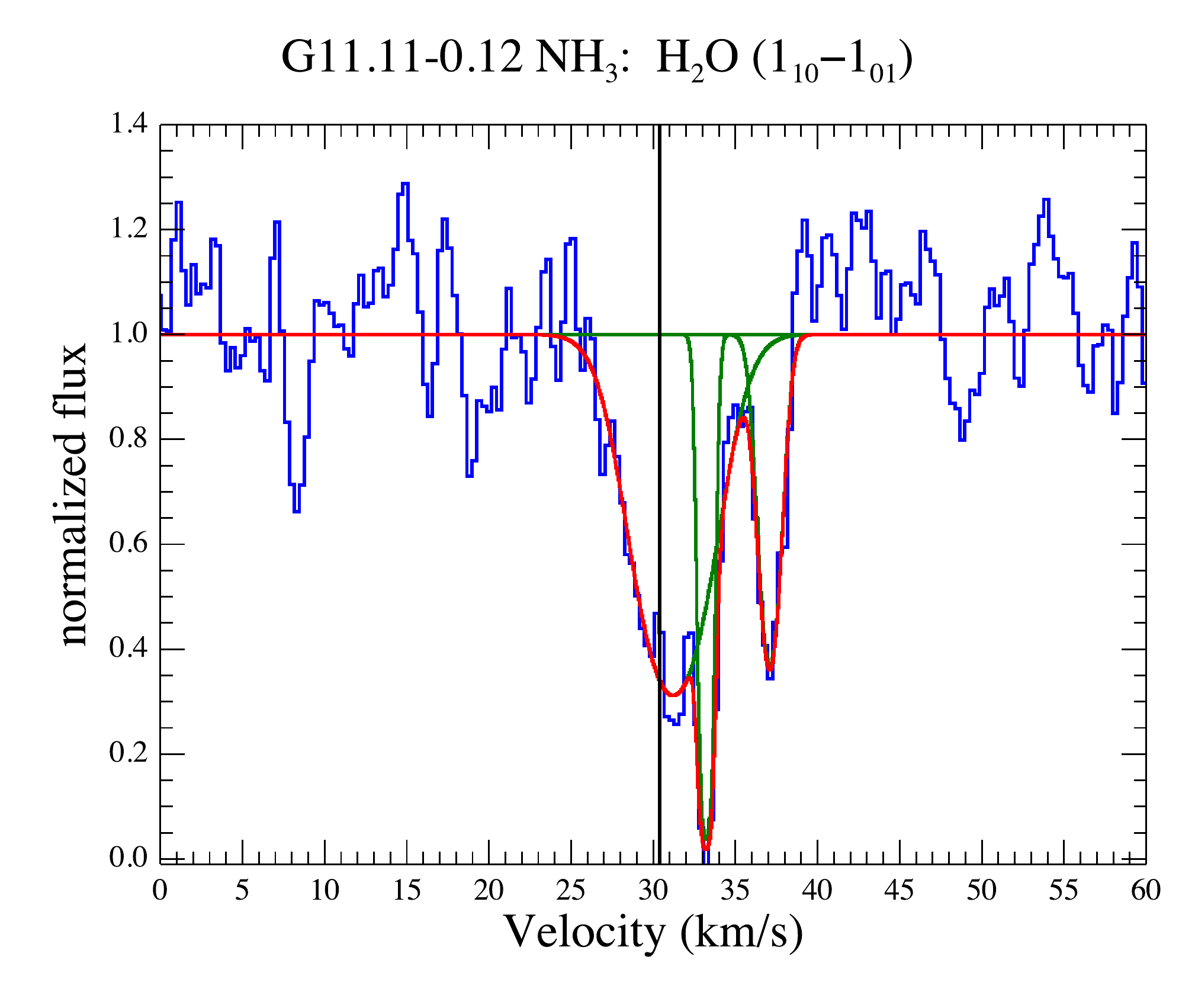}
\end{minipage}
\begin{minipage}{0.45\textwidth}
\centering
\includegraphics[width=6.0cm]{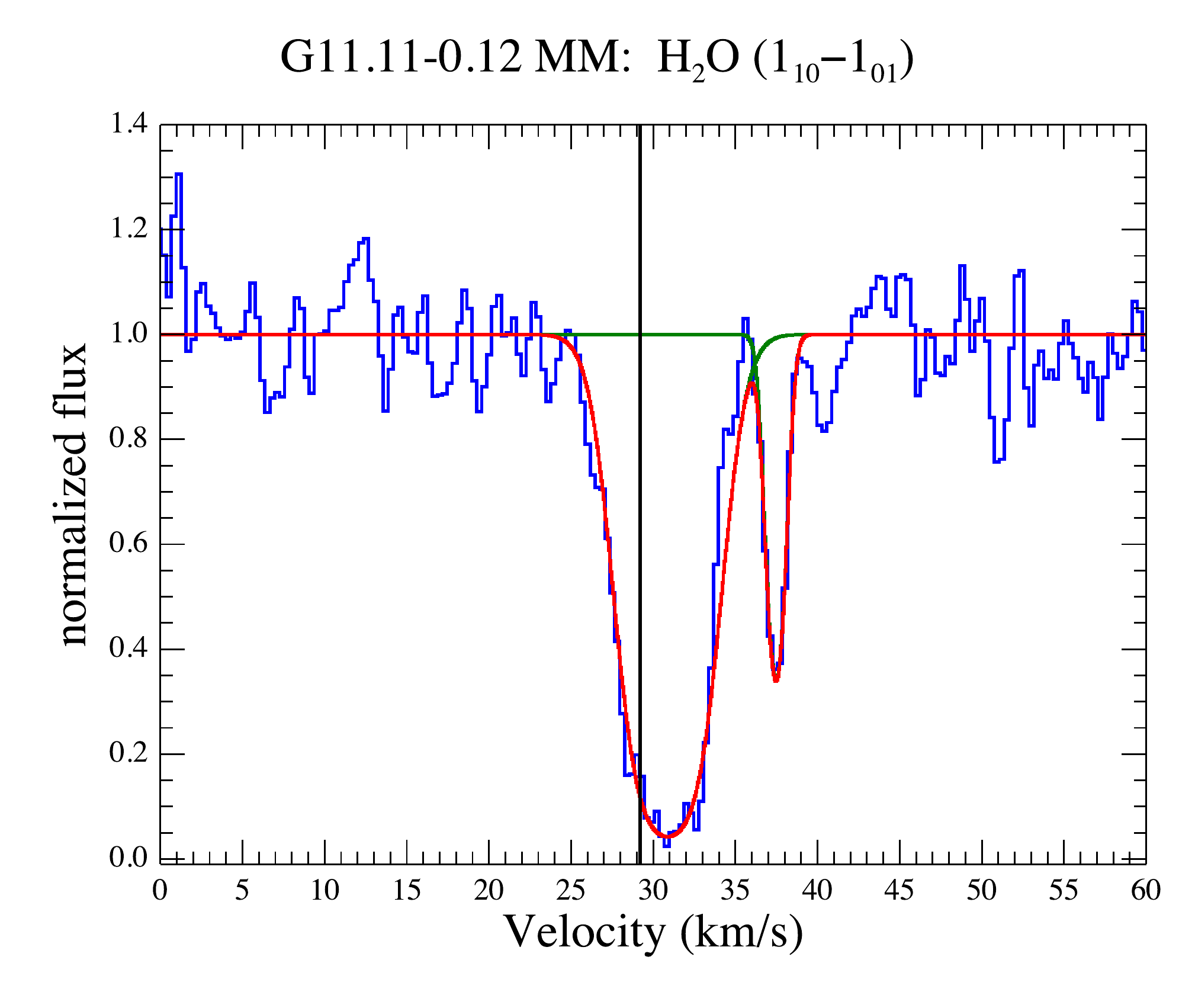}
\end{minipage}
\caption{Main water absorption component in each \bf \rm  clump\rm\label{modelabs}.    The data have been normalized by the total emission models (see Sect. \ref{fittingprocedure}).   The individual models are indicated in green whereas total absorption models are shown in red. }
\end{center}
\end{figure*}

Figure \ref{modelabs} shows the absorption features in relative units, and the results of the fitting procedure are as follows.  
The clump \onea\ is well fit by a single absorbing component at 81.4~\kms ;  
\oneb\ requires at least two components, one broad component at 79.3~\kms and another at 82.8~\kms.  The main absorption model for \oneb\ is not particularly satisfying.  The model underestimates the blue absorption and over estimates the red.   It does however, provide a sufficient representation of the width and depth of the main absorption.   Multiple components for the main feature are possible but are highly degenerate with regards to depth and width without providing more information.  We choose for simplicity to retain only one main absorption feature.    
The clump \twoa\ can be well fit by a broad absorption at 31.2~\kms, a fully saturating feature at 33.2~\kms\ and a narrow feature at 37.1~\kms.   It is 
notable that the broad absorption is not saturated unlike in the other \bf \rm  clumps. 
The clump \twob\ is reasonably well fit by a broad absorption feature at 30.9~\kms\ and a narrow feature at 37.5~\kms.

\begin{table*}
\begin{center}
\caption{Results of Gaussian line fits \label{resultstable}}
\tiny
\begin{tabular}{lcccccll}

\hline \hline
  Line                                    &  Velocity   & $ \Delta v$ &  T$_{mb}$ or $ \tau$\tablefootmark{d} &  Continuum & Noise  \tablefootmark{e}&   notes &\\ 
                                                 &  (\kms)      &    (\kms)   &    (mK)   &       (mK)          & (mK)     &   \\    
\hline
\multicolumn{7}{c}{ \onea} \\
 \hline  

 o-\hto\ 1$_{10} \to 1_{01}$      &  83.9 (0.7)  & 39.5  (2.0)    &  13 (1) &    &  3  &   & \\
 o-\hto\ 1$_{10} \to 1_{01}$      &  79.8 (0.1)  & 9.3  (0.4)    &  36 (2) &    &  3  &   & \\
 o-\hto\ 1$_{10} \to 1_{01}$      &  81.4 (0.1)  &  3.3 (0.1) &  $>$-4.45&  25 (0.5)  &  3  &   abs & \\
 o-\htoiso\ 1$_{10} \to 1_{01}$ &   79.4 (0.1) &  1.2 (0.3)     &  -0.76 (0.15)  &  25 (0.5) &  9&  abs  &   \\
 \ntwohp\ $6 \to 5$                   &  79.8 (0.2) &  2.3 (0.5) & 11 (2) &   &  10 &   \tablefootmark{b} &\\
 \ntwohp\ $3 \to 2$                  &  80.3 (0.1)&  3.4 (0.3)&  827 (68) &   &  250 &  12.5\arcsec \tablefootmark{g} \\ 
 \csto\ $ 3 \to 2 $                      &  79.3 (0.8) &  5.2 (1.9) &  110 (40)   &    &  60  &   \tablefootmark{b} &\\
 \ctfs\ $7 \to 6$                      & ... &  ...  &    ...   &   &  60 &    \tablefootmark{a}  \\
 \meth\  (all lines)     &   ... &   ... &   ...  &   &  60 &  \tablefootmark{a} \\ 

\hline
\multicolumn{7}{c}{ \oneb} \\
 \hline                     
 o-\hto\ 1$_{10} \to 1_{01}$      &  80.5 (0.2)  &   14.4 (0.6)  &  71 (4)  &   &  4 &    & \\
 o-\hto\ 1$_{10} \to 1_{01}$      &  82.1 (0.2)  &   42.9 (0.9)  &  62 (2)  &   &  4 &    & \\


 o-\hto\ 1$_{10} \to 1_{01}$ &   79.3 (0.1)  &   6.6 (0.2)    &  $>$-3.79 &  138 (0.5) &  4&  abs &  \\
 o-\hto\ 1$_{10} \to 1_{01}$ &   82.8 (0.1)  &   1.5 (0.2)    &  -2.16 (0.02) &  138 (0.5) &  4&  abs &  \\
 o-\htoiso\ 1$_{10} \to 1_{01}$ &   78.3 (0.1)  &  1.1 (0.1)     &  -0.41 (0.03) &  129 (0.5)  & 9 &  abs &  \\
 \ntwohp\ $6 \to 5$                   &  78.2 (0.1) &  3.8 (0.1)  &  404 (2) &   &  5 & &  \\
 \ntwohp\ $3 \to 2$                   &  77.8 (0.1) &  5.45 (0.1)  &   3580 (140) &   &  240 &   6.3\arcsec \tablefootmark{g} \\
 \csto\ $ 3 \to 2 $                       &  78.2 (0.1) &  2.9  (0.1)  &  1800 (40)  &    &  60  &  &  \\
 \ctfs\ $7 \to 6$                         & 77.6 (0.1)  &  5.6 (0.2)  &    260 (10)   &   &  60 &   & \\
 \cch\ $4_{55} \to 3_{44}$ 		& 78.5\tablefootmark{c} &  4.20 (0.07) &   680 (10) &  & 60&& \\
 \cch\ $4_{44} \to 3_{33}$ 		& 78.5\tablefootmark{c} &  4.80 (0.09) &   510 (10) &  &  60& &\\
 \meth\  $4_0 \to 3_{-1} -$E         & 78.62 (0.06) & 4.59 (0.19)& 466 (22)   &   &  60 & &\\
  \meth\ $4_0 \to 3_{-1} -$E         & 75.57 (0.42) & 10.27 (0.48) & 186 (18)   &   &  60 & &\\
 \meth\  $7_{-1} \to 6_{-1} -$E      & 78.35 (0.02)& 3.49 (0.07)& 1237 (37)  &  &  60 &  &\\
 \meth\  $7_{-1} \to 6_{-1} -$E      & 78.02 (0.05)& 8.32 (0.22)& 702 (38)  &  &  60 &  &\\

  \meth\ $7_{0 +} \to 6_{0 +} -$A  & 78.32 (0.02)& 4.38 (0.05)& 1917 (23) &    &  60&  & \\
 \meth\  $7_{0 +} \to 6_{0 +} -$A  &  77. 74 (0.01)& 12.12 (0.48)& 313 (23) &    &  60&  & \\

 \meth\ $7_{0} \to 6_{0} -$E          &  78.23 (0.01)& 3.79 (0.03)&  880 (6)  &  & 60 & &   \\
 \meth\ $7_{0} \to 6_{0} -$E          &  78.23 (0.01)& 3.79 (0.03)&  880 (6)  &  & 60 & &   \\
 
 \meth\ $7_{1} \to 6_{1} -$E          &   78.62 (0.02)& 3.90 (0.04)&  492 (5)&    & 60 &  &  \\
 \meth\ $7_{2} \to 6_{2} -$E          &   78.5 \tablefootmark{h} & 4.02 (0.17)&  261 (13)&  &  60 &  &  \\
 \meth\ $7_{-2} \to 6_{-2} -$E       &  78.5 \tablefootmark{h} & 4.30 (0.12)&   355 (13)  &   & 60 &   \\
 \meth\ $7_{-2 -} \to 6_{-2 -} -$A& ... & ... &   $< $50   &   & 60 &   \tablefootmark{a}\\

\hline
\multicolumn{7}{c}{ \twoa} \\
 \hline                     
 o-\hto 1$_{10} \to 1_{01}$      &  31.2 (0.1)  &  4.9 (0.1)   &  -1.17 (0.03) &  35 (0.3) &  4&  abs  & \\
 o-\hto 1$_{10} \to 1_{01}$      &  33.2 (0.1)  &  0.8 (0.1)   &  $>$-3.30  &  35 (0.3) &  4&  abs  & \\
 o-\hto 1$_{10} \to 1_{01}$      &  37.1 (0.1)  &  1.5 (0.1)   &  -1.00 (0.04) &  35 (0.3) &  4 & abs   \tablefootmark{f}\\
 o-\htoiso 1$_{10} \to 1_{01}$ &   29.9 (0.2)  &    2.2 (0.4)   &  -0.51 (0.1) &  31 (0.3) & 8 & abs  & \\
 \ntwohp $6 \to 5$                   & 31.0 (0.2)  &  3.4 (0.4) &  18 (2) & &  4 & &  \\
 \ntwohp $3 \to 2 $                 &  30.7 (0.1) &  2.3 (0.1) &  1080 (40) & &  100 &  22.3\arcsec \tablefootmark{g}  \\
 \csto\ $ 3 \to 2 $                      &  30.7 (0.6)  & 5.1 (1.5)   &  120 (30)   &   &  60  &  &  \\
 \ctfs\ $7 \to 6$                      &  ... &   ... &     ...  &   & 60&   \tablefootmark{a}  \\
 \meth\  (all lines)     &   ... &   ...&  ... &    &  60 &   \tablefootmark{a} \\
\hline
\multicolumn{7}{c}{ \twob} \\
 \hline                     
 o-\hto\ 1$_{10} \to 1_{01}$      & 27.8 (2.0)   &   9.1 (2.3)  & 43 (2) &  &   &  & \\
 o-\hto\ 1$_{10} \to 1_{01}$      & 28.7 (0.2)   &   23.0 (1.3)  &  24 (4) &  &   &  & \\
 o-\hto\ 1$_{10} \to 1_{01}$      & 30.9 (0.1)  &   4.4 (0.1) &  $>$-3.16 &  44 (0.7)  &  5 & abs  & \\

 o-\hto\ 1$_{10} \to 1_{01}$      & 37.5 (0.1)  &   1.2 (0.1)  &  -1.08 (0.05) &  44 (0.7)  &  5 & abs \tablefootmark{f} \\
 o-\htoiso\ 1$_{10} \to 1_{01}$ &   ...  &    ...  & $<$ -0.03  &  40 (0.7)  &  8 &  abs  \tablefootmark{a}  \\
 \ntwohp\ $6 \to 5$                   &  29.6 (0.1)  &  2.7 (0.1)  &  84 (2)&  &  5&  &  \\
 \ntwohp\ $3 \to 2$                   &  29.6 (0.1) &   5.0 (0.1)  &  1800 (43) & &  180 &  0.0\arcsec \tablefootmark{g}  \\
 \csto\ $ 3 \to 2 $                      & 29.6 (0.2)  &  3.0 (0.3)  &  570 (50)   &   &   75 &  &  \\
 \ctfs\ $7 \to 6$                      &  29.2 (0.3) &  6.3 (0.7)  &   160 (10)    &  &  75 &   & \\
 \cch\ $4_{55} \to 3_{44}$   &  29.2\tablefootmark{c} &  4.4 (0.2)& 190 (10) & &  75 &&\\
 \cch\ $4_{44} \to 3_{33}$   &  29.2\tablefootmark{c} &  4.0 (0.2)& 190 (10) & &  75 && \\
 \meth\ $7_{-1} \to 6_{-1} -$E      &  29.44 (0.04)& 5.0 (0.1)&  317 (5) &   & 75 &  & \\
 \meth\ $7_{0} \to 6_{0} -$E      &  29.37 (0.13)& 2.1 (0.3)&   96 (11) &   & 75 &  & \\
 \meth\ $7_{1} \to 6_{1} -$E      &  29.78 (0.09) &3.2 (0.2)&   118 (7) &   & 75 &  & \\
 \meth\ $7_{0 +} \to 6_{0 +} -$A  &   29.9 (0.1)& 4.1 (0.2)&  289 (11)&   & 75 &  & \\
\hline
\end{tabular}

 \tablefoot{
\tablefoottext{a}{Non-detection.}
\tablefoottext{b}{Marginal detection.}
\tablefoottext{c}{Possible blend, systemic velocity assumed.}
\tablefoottext{d}{Negatives indicate optical depth of absorption. }  
\tablefoottext{e}{1$\sigma$ rms.}
\tablefoottext{f}{Foreground.}
\tablefoottext{g}{Pointing offset between \ntwohp\ APEX observations and HIFI}
\tablefoottext{h}{Lines blended. systemic velocity assumed and lines fit simultaneously.}
Uncertainties listed are the formal 1 $\sigma$ errors on the Gaussian fit. Lower limits identified by $>$, upper limits by $<$.
}

\end{center}
\end{table*}


%
%
%
%

\section{Analysis \label{analysis}}

\subsection{Line velocities}
Inspection of Table \ref{resultstable} reveals three different types of molecular lines.  The emission lines all appear at or near the systemic velocity of the \bf \rm  clumps\rm.    The \htoiso\ lines are in absorption at the systemic velocity.   The \hto\ absorption features either appear well offset ($>$ 5 \kms) or at the systemic velocity with a redshifted wing.  The highly offset lines are presented in Sect. \ref{foregroundsection}  while the broad redshifted features will be discussed in more detail in Sect. \ref{infallsection}.

\subsection{Line widths}
The widths of the \hto\ emission range from 9 \kms\ to 43 \kms.  Each \bf \rm  clump \rm displaying emission lines is best fit by two lines, a broad (9 to 14~\kms)  and a very broad (23 to 43~\kms) line.  \hto\ absorption lines are usually  broader than other molecular line transitions for  the \bf \rm  clumps\rm.  When absorption occurs away from  the systemic velocity  (for example \twoa\ and \twob\ absorption near 37 \kms), the line widths are much narrower (1.5 to 2 \kms) and similar to the \bf \rm absorption in foreground clouds \rm  observed by \citet{flagey2013}.   The \htoiso\ line widths are notably narrow ($<2$ \kms) and at the systemic velocity.

\bf \rm  
Transitions of \meth\ are detected in both  \oneb\ and \twob.  In \oneb,  the data quality is sufficient to detect two emission components, one broad and one intermediate for the transitions \meth\ $4_{0} \to 3_{-1} -$E,
\meth\ $7_{-1} \to 6_{-1} -$E, and   \meth\ $7_{0 +} \to 6_{0 +} -$A.  The transitions, \meth\ $7_{2} \to 6_{2} -$E  and \meth\ $7_{-2} \to 6_{-2} -$E are blended.  In this case the central velocity is fixed and both lines are fit simultaneously to give the results in Table \ref{resultstable}.
\rm

Based on Table \ref{resultstable} it is possible to identify 3 width regimes of lines \bf \rm  at the systemic velocity\rm: narrow ($<$3 \kms), intermediate (3-7 \kms) and broad ($>$7 \kms).  Table  \ref{widthsummary} places each transition in one of these categories for each \bf \rm  clump\rm.

\begin{table*}
\caption{Classification of line widths \label{widthsummary}}
\begin{center}
\begin{tabular}{lcccc}
\hline \hline
 Clump &  Narrow ($< 3$\kms) &  Intermediate (3 to 7 \kms) &  Broad ($>7$ \kms) \\
\hline
 \onea&  \htoiso, \ntwohp &  \hto, \csto  &  \hto$_e$ \\
 \oneb&  \hto,\htoiso,\csto  &  \hto,\ntwohp,\meth, \ctfs,\cch & \hto$_e$,\meth \\
 \twoa&  \hto,\htoiso,\ntwohp  &  \hto, \csto & \\
 \twob&  \hto,\ntwohp,\csto &  \hto, \meth, \ctfs, \cch &  \hto$_e$ \\
\hline
\end{tabular}
\tablefoot{\hto\ and \htoiso\ refer to widths of absorption lines.  \hto$_e$ refers to the emission lines.}
\end{center}
\end{table*}

\subsection{Column density from emission lines \label{columndensitysection}}

Table \ref{emissioncolumn} lists the column densities for \hto, \ntwohp, \csto, and \ctfs\ (if present) for the four \bf \rm  clumps\rm.   The column densities are estimated using the \bf \rm  LAMDA \rm  molecular database \citep{schoeier2005} and the RADEX radiative transfer code \citep{vandertak2007} based on the line strengths and widths presented in Table \ref{resultstable}.  Since only one transition is observed for these species, the kinetic temperature observed for \ammonia\ \citep{pillai2006b} is used for
three different gas densities: $10^5, 10^6$ and $10^7$ \ccm.   It is recognized that the application of a uniform density and temperature cannot truly describe the \bf \rm  clumps\rm.   The RADEX results, however, can provide indications of abundances for comparison with other published results.

As can be seen in Table \ref{emissioncolumn}, the \csto\ emission lines are insensitive to the gas density.
The CO column density provides an alternative measure of the total \hh\ column density which has been estimated from dust continuum emission in \citet{pillai2006b}.  
Taking the canonical \csto\  abundance as $4.8 \times 10^{-8}$ \citep{frerken1982},  we derive an \hh\ column density for the four \bf \rm  clumps \rm which is 3 to 7 times lower than the one that is obtained based on dust emission.  This underabundance is similar to that found for low mass pre-stellar cores and attributed to CO depletion \citep{bacmann2002}.

\begin{figure}
\begin{center}
\includegraphics[width=9.0cm]{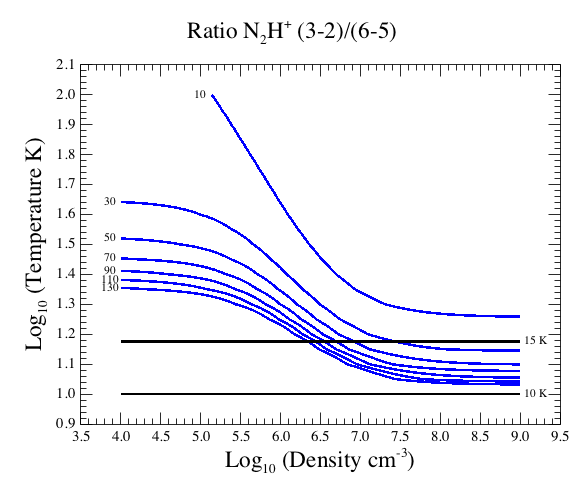}
\caption{Temperature/density trends of the \ntwohp\ (3-2)/(6-5) flux ratio for uniform density, isothermal and optically thin clouds. \bf \rm  Curves of constant flux ratio
are shown in blue.  The flux ratio value of each contour is listed. The black solid line indicates gas temperatures of 10 K and 15 K which are roughly the ranges of the 4 clumps. \rm   \label{lineratiofigure} }
\end{center}
\end{figure}

\bf \rm  An estimation \rm of the gas density is possible based on analysis of the two different transitions of \ntwohp.  The RADEX code is used to determine transition \bf \rm  line flux ratios (where flux=width $\times$ peak intensity) \rm of \ntwohp\ (3-2) to \ntwohp\ (6-5) \bf \rm  for isothermal, uniform density and optically thin clouds\rm.  Figure \ref{lineratiofigure} shows the temperature vs density trends for the \ntwohp\ line flux ratios from 10 to 130.  At a given gas temperature, the figure can be used to narrow the appropriate range of densities.

\bf \rm  The measured flux ratios (and errors) are 110 (26)  for \onea, 13 (1) for \oneb, 41 (5) for \twoa\  and 40 (2) for  \twob.   The temperatures of the clumps determined from \ammonia\ are from 12.7 K to 16 K.  Clump temperatures around 15K would imply densities between $3\times10^6$ \ccm\  to  $3\times10^7$ \ccm.      The clumps would have to be significantly warmer ($>25$ K) than what is found for \ammonia\ to support densities of $10^5$ \ccm\ or lower.   The exception to this is \oneb\ which has a ratio of only 13, perhaps indicating a warmer interior temperature. \rm

\bf \rm 
We note that the \ntwohp\ (3-2) APEX observation is roughly 20\arcsec\ off of the \twoa\ position.   The analysis followed here can only be used to give a general indication of the density of \twoa.  If the flux ratio is a factor of two higher, the resulting density would be about $10^6$ \ccm.  \rm

\bf \rm 
We have explored the effect of increasing optical depth by
increasing the cloud column density in the RADEX calculations by an order of magnitude.  An increase in the optical depth will decrease the density by roughly an order of magnitude for a given temperature and line ratio putting the range of densities for the clumps between $3\times10^5$ \ccm\  to  $3\times10^6$ \ccm .    \rm

These densities are consistent with earlier published results  on \ntwohp\ in \irdcone\ \citep{chen2010} and \ctfs\ in \twob\ \citep{gomez2011}.

\begin{table*}
\begin{center}
\caption{Emission line column densities \label{emissioncolumn}}
\begin{tabular}{lccccccc}

\hline \hline
Line                                    & Velocity   & $\Delta v$ & T$_{mb} $ & N$_5$ & N$_6$&  N$_7$ &\\
                                            & (\kms) &   (\kms)   &   (mK)   & ($10^{14}$~cm$^{-2}$)&($10^{14}$~cm$^{-2}$) &($10^{14}$~cm$^{-2}$)   & \\    
\hline
\multicolumn{7}{c}{ } \\
\multicolumn{7}{c}{\onea\  $T_{kin} $= 13.2 K } \\
 \hline                     
\ntwohp\ $6 \to 5$                   & 79.7 (0.3) & 2.9 (0.7) &11 (2)    & 2.2  & 0.2 & 0.02  & \\
\csto\ $ 3 \to 2 $                      & 79.3 (0.8) & 5.2 (1.9) & 110 (40)  & 5.3  & 4.8  & 4.8 &  \\
\hline
\multicolumn{7}{c}{ } \\
\multicolumn{7}{c}{\oneb\  $T_{kin} $= 16.0 K } \\
 \hline                     
\ntwohp\ $6 \to 5$                   & 78.3 (0.1) & 3.8 (0.1)  & 404 (2) & 27.2 & 2.3 & 0.37&  \\
\csto\ $ 3 \to 2 $                      & 78.2 (0.1) & 2.9  (0.1)  & 1800 (40)  & 40.6   & 37.6 &37.0  &  \\
\ctfs\ $7 \to 6$                         &77.6 (0.1)  & 5.6 (0.2)  &   260 (10)   & 34.8 & 3.0 &  0.38 & \\

\hline
\multicolumn{7}{c}{ } \\
\multicolumn{7}{c}{\twoa\  $T_{kin} $= 12.7 K } \\
 \hline                     

\ntwohp\ $6 \to 5$                   &30.9 (0.1)  & 2.7 (0.3) & 19 (2) & 4.3 & 0.4 & 0.05 &  \\
\csto\ $ 3 \to 2 $                      & 30.7 (0.6)  &5.1 (1.5)   & 120 (30)   & 6.0  & 5.5  & 5.4&  \\
\hline
\multicolumn{7}{c}{ } \\
\multicolumn{7}{c}{\twob\  $T_{kin}$ = 13.8 K } \\
 \hline                     
\ntwohp\ $6 \to 5$                   & 29.6 (0.1)  & 2.7 (0.1)  & 85 (2)& 9.1 &0.92& 0.13 &  \\
\csto\ $ 3 \to 2 $                      &29.6 (0.2)  & 3.0 (0.3)  & 570 (50)   & 15.0  &  14.0& 13.8 &  \\
\ctfs\ $7 \to 6$                      & 29.2 (0.3) & 6.3 (0.7)  &  160 (10)    & 43.0 & 3.8 & 0.46  & \\

\hline
\end{tabular}

\tablefoot{N$_5$ is the calculated column density assuming a gas number density of $10^5$ \ccm, likewise for N$_6$ and N$_7$ }
\end{center}
\end{table*}

\begin{table}
\begin{center}
\caption{Outflow column density for o-\hto  \label{wateremissioncolumn}}
\begin{tabular}{lcccc}

\hline \hline
Line                                    & Velocity   & $\Delta v$ & T$_{mb} $ & N \\
                                            & (\kms) &   (\kms)   &   (mK)   & ($10^{14}$ \scm)  \\    
\hline
\multicolumn{5}{c}{ } \\
\multicolumn{5}{c}{\onea } \\
 \hline                     
o-\hto\     & 83.9 (0.7)  &39.5 (2.0) & 13 (1) &  0.5\\
o-\hto\      & 79.8 (0.1)  &9.3 (0.4) & 36 (2) &  0.3  \\
\hline
\multicolumn{5}{c}{ } \\
\multicolumn{5}{c}{\oneb} \\
 \hline                     
o-\hto\   & 82.1 (0.2)  &  42.9 (0.9)  & 62 (2)  & 3.0  \\
o-\hto\  & 80.5 (0.2)  &  14.4 (0.6)  & 71 (4)  & 1.1  \\

\hline
\multicolumn{5}{c}{ } \\
\multicolumn{5}{c}{\twob } \\
 \hline                     
o-\hto\    &28.7 (0.2)   &  23.0 (1.3)  & 24 (4) & 0.6\\
o-\hto\ &27.8 (2.0)   &  9.1 (2.3)  & 43 (2) & 0.4 \\

\hline
\end{tabular}

\tablefoot{Column density estimate assuming outflow properties of $T_{kin} = 200$ K and $n(H_2) = 3\times10^4$ \ccm from \citet{vandertak2010}. }
\end{center}
\end{table}

\subsubsection{The o-\hto\  emission} 
The \hto\ line profiles are quite different from those of the other lines.  When emission is detected (\onea, \oneb, and \twob),  the line is very broad with line widths of 9 to 45 \kms.  Like for other high mass \hto\ proto-stellar cores \citep[e.g.][]{vandertak2010},  our data are better fit by two broad emission components, a broad component from 9 to 15 \kms\ wide and a very broad component 23 \kms or greater.  As in \citet{vandertak2010} this component is identified as an outflow.   

It is difficult to accurately estimate the column density of \hto\ in the outflow.   The process giving rise to this component is not quiescent and therefore probably at quite different temperatures and densities than the material seen in \ammonia, \ntwohp\ or \ctfs.  Even \csto\ is following the dense core and not tracing the outflow.  
In light of these difficulties, we follow the approach of \citet{vandertak2010} and calculate the column density in the outflows assuming that  $T_{kin} =200$ K and  $n=3\times 10^4$ \ccm .  These results are shown in Table \ref{wateremissioncolumn}.  

\subsection{Column density from o-\hto\ absorption}

The column density of the absorbing water follows readily from the depth of the line and its width \citep{plume2004}.
Table \ref{waterabsorptiontable}  shows the column density and abundance of water based on absorption measurements.  In some cases, the values listed are lower limits to the column density, since the water absorption lines are close to saturation.

Table \ref{waterabsorptiontable} also shows the results from \htoiso.  The isotope provides an independent measurement of the water column density when present or an upper limit when it is not detected.   For the upper limits, the depth of the line is estimated as 1$\sigma$ of the single sideband continuum with a line width of 2 \kms.   \bf \rm  The \htoiso\ absorption features are narrow (1-2 \kms) and at the systemic velocity while the main \hto\  features are mostly saturated,  broad (3-6 \kms) and redshifted by $\sim 2$ \kms (see \ref{infallsection}).  This discrepancy implies that part of the absorption seen in \hto\ is a different component that what is probed by \htoiso.  The abundance listed in Table \ref{waterabsorptiontable} is only calculated based on the \htoiso\ data.


Both \bf \rm  clumps \rm of \irdcone\ display an \hto\ absorption complex between 0 and 30 \kms\ LSR.  These features are shown in Fig. \ref{irdconeforeground} and discussed in Sect. \ref{foregroundsection} below.

\begin{table*}
\begin{center}
\caption{o-\hto\ Column density from absorption lines \label{waterabsorptiontable}}
\begin{tabular}{rlccccc}
\hline\hline
ID  & $v$ & $\Delta v$    &$\tau$& N(o-\hto) & $\chi$  \\
    & (\kms) & (\kms)   &  & ($10^{14} cm^{-2})$ & ($10^{-8}$)   \\

\hline
 &&&&&\\

\onea & 81.4                                  &   3.3    & $>4.5$\tablefootmark{a}& $>0.72$ &  ...     \\
           &79.4\tablefootmark{c}       & 1.2    &    0.76                               &  21.6     & 3.7   \\
\hline
 &&&&&\\

\oneb & 79.3                                  &   6.6 & $ >3.8$\tablefootmark{a} & $>1.2$&  ...       \\
          & 82.8                                  &   1.5 & $ 2.2$\tablefootmark{a} & $0.16$&     ...    \\
	 &  78.3\tablefootmark{c} 	 & 1.1   &      0.4                             &   10.8    & 0.33\\
\hline
 &&&&&\\

\twoa & 31.2  			    &  4.9    & 1.7                                 & 0.28     & 	...	 \\
	& 33.2                               &   0.76   & $>3.3$\tablefootmark{a} & $>0.12$&    ...     \\
           & 29.9\tablefootmark{c}    &  2.2    & 0.5                & 26.1    &  	3.1	 \\
\hline
 &&&&&\\
\twob & 30.9                                & 4.4     & $>3.2$\tablefootmark{a}   & $>0.69$ & ...  \\
          & 30.0\tablefootmark{b}  & 2.0     & $<0.03$   & $<1.5$            &   $0.07-0.15$ \\
\hline
\end{tabular}
\tablefoot{
\tablefoottext{a}{Lower limit, since line is very close to saturated}
\tablefoottext{b}{Estimated upper limit based on non detection of \htoiso\ feature assuming an abundance with respect to \hto\ of 500 and adopting a line width of 2.0 \kms.}
\tablefoottext{c}{Measured \htoiso\ and abundance with respect to \hto\ of 500}
}
\end{center}
\end{table*}


\subsection{The \meth\ -E excitation analysis}
\begin{figure}
\includegraphics[width=9.0cm]{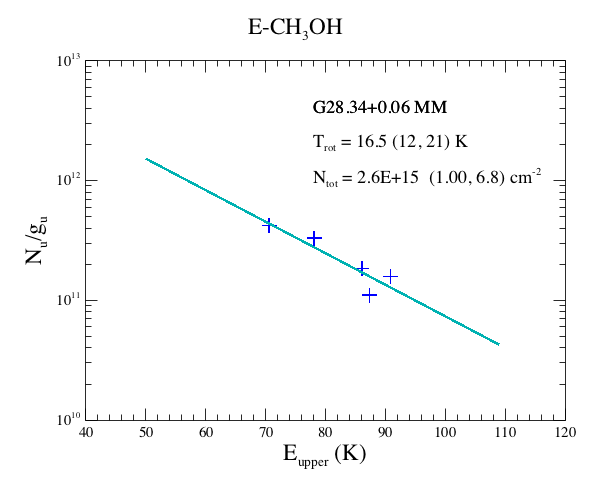}
\caption{The E-methanol rotation diagram for \oneb.  \label{rotation_oneb} }
\end{figure}

\begin{figure}
\includegraphics[width=9.0cm]{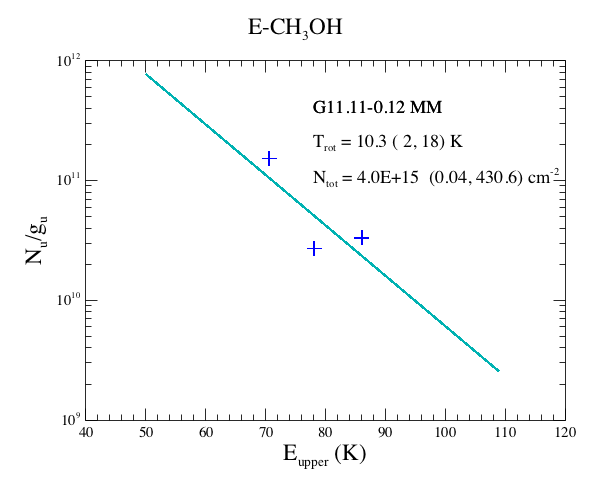}
\caption{The E-methanol rotation diagram for \twob.  \label{rotation_onea} }
\end{figure}

In the \bf \rm  clumps \rm \oneb\ and \twob\, \meth\ transitions are observed.   
Using the formalism of \citet{helmich1994} and \citet{blake1987}, rotation diagrams of the \meth\ transitions can be calculated.  The line strength and permanent dipole parameters are taken from \citet{anderson1990}.  The resulting rotation diagrams are shown in Figs. \ref{rotation_oneb} and \ref{rotation_onea}.   \bf \rm  This analysis assumes optically thin transitions in LTE.   We use only the narrower line component of \oneb . The results indicate a rotation temperature of 10 K to 16 K and a column density of \meth\ of  4.0 to 2.6$\times10^{15}$ \cms.
With only three data points for \twob, the column density of E-\meth\ is poorly constrained. \rm
    
\bf \rm 
\citet[][]{leurini2007} observed 5 IRDC clumps in \meth\ transitions to probe their density and temperature structure. In general they find that clumps need upto  three components:  an extended component, a core component and an outflow component.  One of their clumps is \twob.  They conclude that two components are needed to describe the emission:  extended and core.   The kinematic temperatures they derive are $16-19$ K (extended) and $38-54$ K (core)  They did not observe the 338 GHz band toward this source.    The rotation temperature and column density we find is most in agreement with their extended component.  
Although \oneb\ was not observed  by \citet[][]{leurini2007}, the broad \meth\ lines that are found here would be classified as an outflow component. \rm


\begin{figure}
\includegraphics[width=9.0cm]{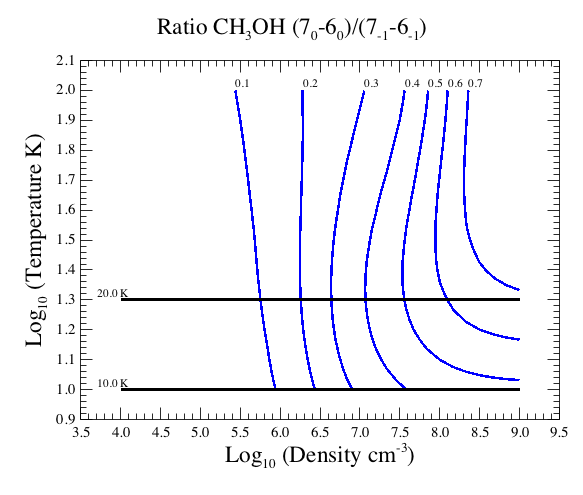}
\caption{The \meth -E ratio plot for densities from 10$^4$ \ccm\ to 10$^9$ \ccm\ and gas temperatures from 10 K to 100 K.   Curves of constant flux ratio
are shown in blue.  The flux ratio value is listed above these curves. The black solid line indicates gas temperatures of 10 K and 20 K which are roughly the kinematic temperatures of the rotation analysis.\label{methratio} }
\end{figure}

In order to estimate the density of the gas from which the \meth\ lines originate,  
we perform a similar  analysis for ratios of \meth\ lines as for the ratio of \ntwohp\ transitions.  Choosing the two brightest \meth\ transitions $7_{-1} - 6_{-1}$ and $7_{0} - 6_{0}$,  line ratios of  0.77 for \oneb\ and 0.13 for \twob\ are obtained.   We calculate a grid of RADEX models for densities from 10$^4$ \ccm\ to 10$^9$ \ccm\ and gas temperatures from 10 K to 100 K.    From Fig. \ref{methratio} we can see that the ratio of the transitions is relatively insensitive to the gas temperature and that for the observed ratio \bf \rm  \rm for \twob\   a density of $3\times 10^6$ \ccm\ is suggested.   For \oneb\ , the ratio is quite high which could be indicating that \meth\ is originating from a higher temperature component.   Nevertheless the analysis is indicating high densities $> 10^7$ \ccm.

\section{Discussion \label{discussion}}

\subsection{Foreground absorption \label{foregroundsection}}
Each cloud shows absorption features at velocities 5 \kms\ or more different from the systemic velocity of the cloud.  These features are usually narrow and not fully saturated.  Examples of such clouds are found in \citet{marseille2010} and \citet{flagey2013}.   These are taken to be foreground clouds. 

\begin{figure}
\includegraphics[width=8.0cm]{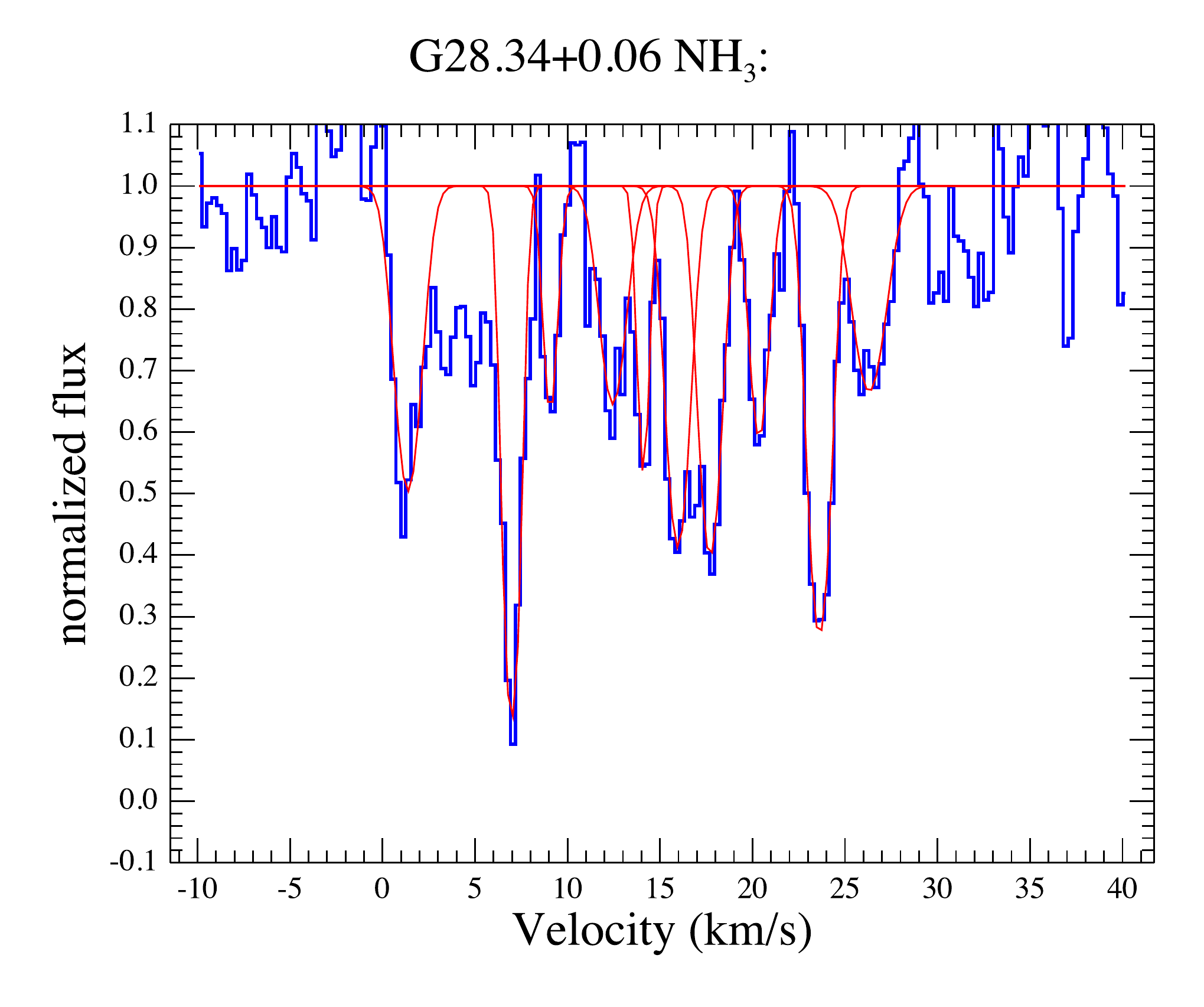}
\includegraphics[width=8.0cm]{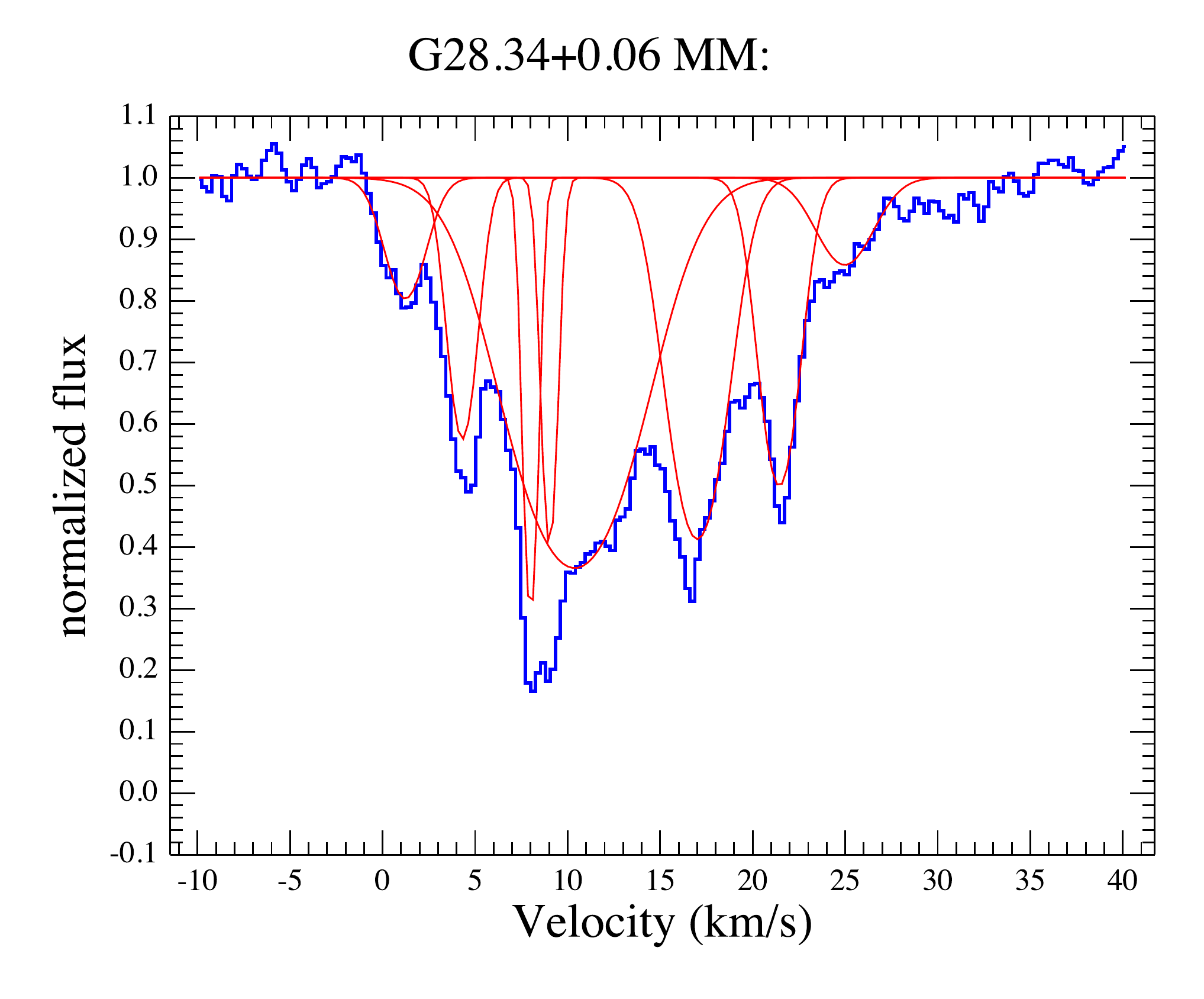}
\caption{Absorption for \onea\ and \oneb\ between 0 and 30 \kms LSR.  The data in blue are normalized single sideband flux.   Identified features are shown in red.
}\label{irdconeforeground}
\end{figure}

Toward \onea\ and \oneb\ there is a complex of water absorption features between 0 and 30 \kms\ LSR (see Fig. \ref{irdconeforeground}).    Column densities for these clouds are estimated to be between $1\times10^{12}$~cm$^{-2}$  and $4\times10^{13}$~cm$^{-2}$.  Both absorption complexes show roughly similar features (deep narrow absorption around 7 to 9~\kms,  another set around 15~\kms and again near 22~\kms).   These similar features suggest a common cloud in front of the two \bf \rm  clumps \rm with possibly a velocity gradient of a few \bf \rm   \kms\  \rm between the two \bf \rm  clump \rm positions.

In both \bf \rm  clumps \rm \twoa\ and \twob\ there are features near 37 \kms .  The difference in velocity from systemic as well as the similar width and column densities of the features, suggests that they also originate from foreground clouds.    Table \ref{fgabsorptiontable} lists the velocities, widths and column densities for all features identified as foreground for o-\hto. \bf \rm  The widths and column densities are very similar to the foreground clouds towards W51 \citep[][]{flagey2013}. \rm

\begin{table}
\begin{center}
\caption{Foreground o-\hto\ absorption. \label{fgabsorptiontable}}
\begin{tabular}{rcccc}
\hline\hline
ID  & $v$ & $\Delta v$    &$\tau$& N(\hto)   \\
    & (\kms) & (\kms)   &  & ($10^{14} cm^{-2})$  \\
\hline

\onea    & 1.4 & 1.6 & 0.67  & 0.05 \\
& 7.0  & 0.9 & 1.91  & 0.09 \\
& 9.1 & 0.9 & 0.45  & 0.02 \\
& 12.5 & 1.6 & 0.43  & 0.03 \\
& 14.1 & 0.7 & 0.62  & 0.02 \\
& 16.0  & 1.4 & 0.87  & 0.06 \\
& 17.7 & 1.4 & 0.91  & 0.06 \\
& 20.4 & 1.2 & 0.52  & 0.03 \\
& 23.6 & 1.3 & 1.30   & 0.08 \\
& 26.3 & 2.0 & 0.41  & 0.04 \\


\oneb  & 1.2 & 2.5 & 0.24  &0.03  \\
& 4.3 & 1.8 & 0.62  & 0.05  \\
& 8.0 & 0.8 & 1.44  & 0.06  \\
& 9.0 & 0.9 & 1.04  & 0.05  \\
& 10.4 & 7.3 & 1.15  &  0.41  \\
& 17.0 & 3.4 & 0.98  & 0.16   \\
& 21.4 & 2.4 & 0.75 &  0.10    \\
& 25.0 & 3.5 & 0.16  &  0.03  \\

\twoa  & 37.1   & 1.5    & 1.0    & 0.07 \\
\twob &  37.5  & 1.2     & 1.1 & 0.07  \\
\hline
\end{tabular}
\tablefoot{Table of column density estimates of  absorption features at significantly different velocities than the \bf \rm  clumps\rm.  Features are shown in Fig. \ref{irdconeforeground} }
\end{center}
\end{table}

\subsection{Infall \label{infallsection}}
The \hto\ absorption is systematically redshifted relative to the systemic velocity of each \bf \rm  clump\rm.  Since this absorption is seen against the dust continuum emission or the very broad \hto\  emission, this motion is detected between the observer and the source of the emission. The absorbing water is moving towards the IRDC \bf \rm  clumps\rm.   The redshifted absorption is not seen in the \htoiso\ absorption lines nor in the \ntwohp, \csto, or \meth\ emission.   The completely saturated absorption indicates that the infall is occurring on spatial scales at least as large as the extent of the continuum emission.

If interpreted as gas falling toward a central point, we can attempt to estimate the infalling material rate \citep{beltran2006}.   The rate is simply the velocity of gas, $V_{infall}$, at a given distance  from the centre, $R$ , the density of this gas, $n(R)$ and the molecular weight $\mu$.
\[  \dot{M} = 4 \pi R^2 \mu_{H_2} n(R) V_{infall}, \]
assuming that the material is falling from every direction toward the centre of the \bf \rm  clump \rm and gives an upper limit to the rate if the infall is not uniform.
Three of the four \bf \rm  clumps \rm have been observed at high spatial resolution.  \bf \rm  \onea\ and \oneb\ in \ntwohp\ (3-2) and continuum \citep[][]{chen2010} and \twob\ in \meth\ and \ctfs\ \citep[][]{gomez2011}. \rm  None of the high resolution observations identify infalling gas.   The \htoiso\ lines are not redshifted and are therefore more associated with the \ctfs, \ntwohp\ gas.
This indicates, therefore, that infalling gas is occurring at larger radii than these observations.  

We can then use the sizes determined by the high angular resolution observations to provide an upper limit to the infall rate.  Table \ref{infalltable}  lists these values under the assumption of three different densities.  If the \bf \rm  clumps \rm have a power law density structure with an index of $-2$, the infall rate should be roughly constant.  If this is the case and given the gas density determined by the \ntwohp\ and \meth\ ratios is estimated to be on the order of $10^6$ \cmc,  our best estimate of the infall rate is $\sim 10^{-3}$ \Msun\ yr$^{-1}$.

\bf \rm 
Power laws with index from -1 to -2.5 have been seen observationally and depend on distance from the central source \citep[][]{vandertak2000, beuther2002}.    The adoption of a power law of $-2$ would significantly over estimate the infall rate at large distances and underestimate the rate at close distances.  Clearly a physical structure for the clumps model is needed.  
\rm

\begin{table*} 
\center
\caption{Mass infall rates \label{infalltable}}

\begin{tabular}{lccccc}
\hline 
\hline

\bf \rm  Clump\rm&V$_{inf}$ &R &  $\dot{M}_5$\tablefootmark{a} & $\dot{M}_6$ & $\dot{M}_7$ \\
       & (\kms)      & (pc)  & (\Msun\ yr$^{-1}$) & (\Msun\ yr$^{-1}$)  & (\Msun\ yr$^{-1}$) \\
\hline
\onea & 1.2 & 0.05\tablefootmark{1} & 2.1e-04 & 2.1e-03 & 2.1e-02 \\
\oneb & 0.8 & 0.1\tablefootmark{1} & 5.7e-04 & 5.7e-03 & 5.7e-02 \\
\twoa & 0.8 & 0.04 \tablefootmark{b} & 9.2e-05 & 9.2e-04 & 9.2e-03 \\
\twob  & 1.7 & 0.04\tablefootmark{2}& 1.9e-04 & 1.9e-03 & 1.9e-02 \\
\hline
 \end{tabular}
 \tablefoot{
 \tablefoottext{a}{Infall rate for velocity of broadest absorption component assuming densities of $10^{5}$\ccm, $10^{6}$\ccm\ and $10^{7}$\ccm }
 \tablefoottext{b}{Core size assumed to be the same \twob.}
References. (1) \citet{chen2010}; (2) \citet{gomez2011}
}
 \end{table*}

%



\subsection{Structure of IRDC  clumps \label{structure}}

  
Our observations reveal different dynamical structures.
\bf \rm  The very broad water lines are readily attributed to molecular outflows.   These are a relatively common feature in \hto\ observations of starforming regions \cite[][]{vandertak2013, kristensen2012}.   Outflows are seen toward three of the four clumps.  A further inspection of the \meth\ data for \oneb\ reveals that here too is an outflow in \meth.
\rm

\bf \rm  In our sample, intermediate line widths (3-7 \kms) are observed in the more  complex molecules.  This component is most akin to the envelope  reported by \citet{vandertak2013} and possibly the extended component noted by \citet[][]{leurini2007}.   We will call this the envelope.
\rm

\bf \rm  The narrow emission lines of \csto\ and \ntwohp\ and \htoiso\ absorption  are consistent with a more quiescent component most akin to the IRDC filament itself \citep[][]{pillai2006b, henshaw2013}. We will call this the quiescent outer envelope.  It is the stage between the gas and dust swept up into the forming proto-star and the molecular cloud filament (IRDC)  \rm

Table \ref{structuretable} lays out the various \bf \rm  clump \rm components, the properties which distinguish them and the tracers that were used to identify each component.





The \hto\ absorption features have intermediate widths, but in all of the \bf \rm  clumps \rm in our sample the \hto\ is redshifted and the absorption is close to completely saturated.   It is not obvious where this layer is to be placed relative to the filament and envelope components. There are three possibilities for where this collapsing absorption layer is occurring.  
\begin{itemize}
\item 
Since the line width of the collapsing \hto\ is similar to the line width of the more complex molecules like \meth, perhaps the layer is interior to the "envelope" region.   This would be a warm and dense infalling gas.   
It is difficult to imagine that the absorbing layer spatially covers the entire emitting region to produce saturated absorption lines.   
\item 
Alternatively, the layer could be between the quiescent outer envelope and the envelope.     In this scenario, the central part of the quiescent outer envelope breaks off to fall onto the envelope. \citet[][]{mottram2013} observed infall from the material surrounding the envelopes of low-mass class 0/I objects.   This layer would be intermediate in density between the \bf \rm  quiescent outer envelope and the envelope\rm.  redshifted wings of other dense tracers like \ammonia, \csto,  and \ntwohp\ are not seen.  Detailed modelling is needed to assess whether this scenario could reproduce the broad saturated lines. 
\item 
\bf \rm  A final possibility is that the absorbing layer is due to gas falling onto the quiescent outer envelope from the filament itself.   In this scenario, the infalling material is essentially falling onto the clump. \rm 
Infalling molecular gas onto filaments has been observed onto the DR21 ridge \citep{schneider2010}.   Since IRDCs are also largely filamentary in nature, it should not be surprising to observe infalling material on IRDC \bf \rm  clumps\rm.   Presumably, the entire IRDC complex  would have shown infalling \hto\ had it been observed.   Since the outer layers of the envelopes have significantly lower line widths than the infalling \hto,  the infalling gas should be relatively low density not to disturb the envelope.  This scenario puts the absorbing layer the farthest from the core and dust emission, making it easier to produce saturated absorption.   This possibility also provides a natural explanation for the lack of  outflow in \twoa.  The collapsing gas seen through the \hto\ absorption is not directly linked with the outflow. 
\end{itemize}

 \begin{table*}
\caption{Dynamic structures within IRDC \bf \rm  clumps\rm  \label{structuretable}}
\begin{center}
\begin{tabular}{lcccc}
\hline
\hline
Component & Properites& Tracer &  \\
\hline
Quiescent envelope& $\Delta$V $< 3$ \kms\  at systemic V& \htoiso, \ntwohp, \ammonia \tablefootmark{a}, \csto \tablefootmark{b}&  \\
Envelope  & $\Delta$V from 3 to 7 \kms\ at systemic V  &  \meth, \ctfs, \cch, \csto \tablefootmark{b} &  \\
Infall& $\Delta$V from 3 to 7 \kms\ redshifted V & \hto\ absorption & \\
Outflow&$\Delta$V $>7$ \kms at systemic velocity& \hto\ emission &  \\
\hline
\end{tabular}
\tablefoot{
\tablefoottext{a}{From \citet{pillai2006b}}.
\tablefoottext{b}{\csto\ is broader than 3 \kms\ for the \ammonia\ positions.}
}
\end{center}
\end{table*}



\subsection{Water in IRDCs}

The IRDCs share many similarities with both low mass and high mass proto-stellar objects.  
The broad \hto\ emission in IRDCs is consistent with observations of high mass and low mass proto-stellar objects \citep{vandertak2013, kristensen2012}.    Unfortunately without a molecular tracer of the outflowing gas, it is not possible to estimate the abundance of \hto\ in the outflow.  Further $^{12}$CO observations should provide the necessary information to pin down the abundance within the outflow.

The o-\hto\ absorption features found here are broader than the low mass counterparts \bf \rm  \citep{kristensen2012,mottram2013} \rm but comparable to the intermediate width features seen in high-mass proto-stetllar objects \citep{vandertak2013}.   The narrow features of \htoiso\ and \ntwohp\ are reminiscent of similar features seen towards low mass envelopes.   

The fact that \htoiso\ is not redshifted indicates that there is a quiescent component in the envelope that is not yet participating in any accretion.   In another high mass proto-star, W43-MM1,  the centres of the \htoiso\ lines are well aligned with the systemic velocity of the clouds but significantly broader than in this work, while the main isotope lines are redshifted in absorption \citep{herpin2012}.   The source  W43-MM1 also shows \hto\ absorption on top of broad emission.  

In Table \ref{waterabsorptiontable}, we estimated the abundance of water vapour.  This calculation assumes that all the \hh\ column density as obtained by the dust emission is involved in the collapse.  The  envelope and core components indicate that parts of \bf \rm  clump \rm that are not collapsing.   Only a fraction of the \hh\ column density is involved in the collapse, which makes our estimate of the water vapour abundance a lower limit.    To obtain a better estimate of the abundance, observations of an independent tracer of the collapsing material (e.g. from $^{12}$CO or $^{13}$CO) are needed.

The water abundance determined from the absorption is consistent with the abundance determined from the low mass source L1544 \citep{caselli2012}.  This level is more indicative of cosmic-ray desorption from grain mantles than thermal evaporation from hot dust.  In low mass objects this absorption occurs in a surrounding envelope, either infalling or expanding.  The \bf \rm  clumps \rm in this work generally only display redshifted infalling motions.

Of the four \bf \rm  clumps\rm, \twoa\ stands out the most in terms of the lack of molecular complexity and outflow activity.  
In a recent study of IRDC G11.11-0.12, \citet{henning2010}  detected 24 far-infrared (70-140 \um) sources and were able to further constrain the emission properties of \twob.   In their work,  a source near  \twoa\ is identified, but close inspection places their source no. 8 more than 20\arcsec\ distant from \twoa\ \citep[][]{chavarria2014};  \bf \rm  \twoa\ has no far-infrared  counterpart and is therefore a high-mass pre-stellar core. 
\rm

\section{Conclusions \label{conclusions}}
The main findings in this study are:
\begin{itemize}
\item
For IRDC \bf \rm  clumps\rm, \hto\ absorption indicates a low gas phase abundance and is more consistent with desorption of \hto\ from dust grains than by heating, similar to low-mass pre-stellar cores.  
\item 
Measurements of \hto\ together with other molecules provide a picture of the structure of massive \bf \rm  clumps \rm on the scale of  about 1 pc, from \bf \rm  outflow, to envelope and quiescent outer envelope, to infalling gas. \rm
\item
Analysis of \ntwohp\ and \meth\ line ratios favour a high internal density ($\sim 10^7$ \ccm).  Further detailed modelling of each \bf \rm  clump \rm will be necessary to determine a proper structure model. 
\item
Molecular \bf \rm  clumps \rm within IRDCs are actively forming stars or star clusters, but at different evolutionary stages.  \ammonia\ observations help to identify the earliest stages, but \hto\ reveals the infall and outflow dynamics \bf \rm  within the  clump\rm.  Strong far-infrared and \meth\  emission as well as chemical complexity are signposts for a \bf \rm  starforming clump \rm in a later stage of evolution.
\item
Even with significant infall motions, \twoa\ does not display outflow motions nor chemical complexity which are signposts of starformation.  This \bf \rm  clump \rm appears to be the only starless massive \bf \rm  core \rm in our sample.
\end{itemize}

Based on a detailed analysis of outflow and molecular complexity it is possible to place the \bf \rm  clumps \rm in a evolutionary sequence.  

\begin{enumerate}
\item 
The clump \oneb\ appears to be the most advanced: it is a \bf \rm bright far-infrared source \rm displaying molecular complexity ("hot core"), \hto\ and \meth\ masers, outflows in both \hto\ and \meth\ lines.   
\item
The clump \twob\ is also an evolved proto-star,\bf \rm also an identified far-infared source \rm with \meth\ masers \bf \rm  and outflows in \hto\ and \meth. \rm  
\item
The clump \onea\ is not as chemically complex as the \submm\ positions \bf \rm  and is \bf \rm not a notable far-infrared source \rm but does display an outflow.
\item
The clump \twoa\ is the least evolved.  \bf \rm It is also not identified in the far-infrared and it does not show \rm out flowing gas,  complex molecules nor maser emission.  
\end{enumerate}

\begin{acknowledgements}
\bf \rm 
We would like to thank L. Pagani, J. Mottram and E. van Dishoeck for careful readings of this manuscript.  We also thank an anonymous referee for thoughtful comments and useful suggestions which significantly improved the final manuscript.  HIFI has been designed and built by a consortium of institutes and university departments from across Europe, Canada and the United States under the leadership of SRON Netherlands Institute for Space Research, Groningen, The Netherlands and with major contributions from Germany, France and the US. Consortium members are: Canada: CSA, U.Waterloo; France: CESR, LAB, LERMA, IRAM; Germany: KOSMA, MPIfR, MPS; Ireland, NUI Maynooth; Italy: ASI, IFSI-INAF, Osservatorio Astrofisico di Arcetri-INAF; Netherlands: SRON, TUD; Poland: CAMK, CBK; Spain: Observatorio Astronómico Nacional (IGN), Centro de Astrobiología (CSIC-INTA). Sweden: Chalmers University of Technology - MC2, RSS \& GARD; Onsala Space Observatory; Swedish National Space Board, Stockholm University - Stockholm Observatory; Switzerland: ETH Zurich, FHNW; USA: Caltech, JPL, NHSC.
\rm
\end{acknowledgements}

\bibliography{./irdcs.bib} 
\bibliographystyle{aa}
\appendix
\section{Sideband calibration \label{sidebandcorrection}}
Since the data show both emission and deep absorption, special care must be taken to handle the double sideband nature of the HIFI data.   The term "source"  sideband refers to the sideband which contains the frequency of the transition of interest.  The "image" sideband refers to the other sideband which is combined in the receiver and ends up in the intermediate frequency.   Line absorption occurs against dust and line emission in the source sideband.  The double sideband receiver therefore adds line free emission from the image sideband.   To correctly calibrate the flux in the source sideband, the continuum flux from the image sideband must be removed.   The continuum emission is due to the dust emission and can differ by a few percent between the source and image sidebands which are separated by about 12~GHz.  Furthermore,  the HIFI mixers in band 1A, are slightly imbalanced also by about a few percent.  In the case of the 557 line in band 1A, all the differences between sidebands can result in about a 5\% error.

Assuming there is no line emission from the image sideband,   a separation of source and image  sidebands is possible using  the latest sideband ratio determination for the 557 line in band 1A \citep{teyssier2013} and an estimate of the ratio of the dust emission emissivity between the source and image sidebands.  For the later, we used the dust grain models of \citet{ossenkopf1994} with thick mantles at 10$^6~cm^{-3}$.  This is the same grain model and dust temperatures used by \citet{pillai2006a} to determine the dust mass.   

The formalism of \citet{flagey2013} is used to correct for the continuum emission from the image sideband.   Each sideband receives a fraction of the emission determined by the sideband gain $\gamma_{upper}$ or $\gamma_{lower}$.  The total continuum is then,
\[
I_{cont} = \gamma_u I_{u,dust} + \gamma_l I_{l,dust} ,
\]
where,  
\[ \gamma_{u} + \gamma_{l} = 1.0, \]
for normalized gains.

Since a dust model and temperature  are assumed, the relative contributions from the dust emission is known and a pseudo continuum from the image sideband can be removed.   This was done for the \hto\ and \htoiso\ data shown in Fig. \ref{observedlineplots}.  These figures show the resulting single sideband data.

\end{document}